\documentclass[12pt,letterpaper]{article}
\usepackage[a4paper, total={7in, 10in}]{geometry}

\usepackage{graphicx}
\usepackage{helvet}
\usepackage{authblk}
\usepackage{hyperref}
\usepackage{amsmath} 
\usepackage{amssymb} 
\usepackage{orcidlink} 
\usepackage[super,comma,sort&compress]  
   {natbib}

\makeatletter
\renewcommand{\maketitle}{\bgroup\setlength{\parindent}{0pt}
\begin{flushleft}
  \textbf{\@title}
  
  \@author
\end{flushleft}\egroup}
\makeatother

\title{Electrochemical reactions under reverse bias create additional mobile ions that enable hole tunneling in metal halide perovskite diodes}
\date{\today}

\author[1,2,9,\orcidlink{0009-0009-1184-7956}]{Kell Fremouw}
\author[2,9,\orcidlink{0000-0003-3219-7736}]{Ryan A. DeCrescent}
\author[3]{Xianfu Zhang}
\author[3]{Yi Yang}
\author[3]{Bin Chen}
\author[1,2]{Daniel A. Morales Jr}
\author[1,2,\orcidlink{0009-0001-9714-8778}]{Matteo R. S. Poma}
\author[4]{Kelly Schutt}
\author[5]{Fangyuan Jiang}
\author[3]{Edward H. Sargent}
\author[6]{Neal R. Armstrong}
\author[5]{David S. Ginger}
\author[2,4]{Joseph M. Luther}
\author[1,2,7,8,10,11,*,\orcidlink{0000-0001-9609-9030 }]{Michael D. McGehee}

\affil[1]{Materials Science and Engineering, University of Colorado, Boulder, CO 80303, United States}
\affil[2]{RASEI, University of Colorado, Boulder, CO 80303, United States}
\affil[3]{Department of Chemistry, Northwestern University, Evanston, IL 60208, United States}
\affil[4]{National Laboratory of the Rockies, Golden, CO 80401}
\affil[5]{Department of Chemistry, University of Washington, Seattle, WA 98195, United States}
\affil[6]{Department of Chemistry and Biochemistry, The University of Arizona, Tucson, Arizona 85721, United States}
\affil[7]{Department of Chemistry and Biochemistry, University of Colorado, Boulder, CO 80303, United States}
\affil[8]{Chemical and Biological Engineering, University of Colorado, Boulder, CO 80303, United States}
\affil[9]{These authors contributed equally}
\affil[10]{Senior author}
\affil[11]{Lead contact}

\affil[*]{Correspondence: michael.mcgehee@colorado.edu}

\begin{document}

\maketitle

\section*{SUMMARY}
Gradual reverse-bias breakdown in metal-halide perovskite diodes and solar cells is thought to originate from hole tunneling through steep bands in an ionic depletion region near the electron transport layer after positively charged iodine vacancies accumulate near the hole-transport layer (HTL). However, typical reported mobile ion concentrations near $1\times10^{17}$ cm$^{-3}$ are too small to quantitatively explain significant tunneling current densities and (Zener) breakdown observed near $-5$ V. Here, we show that inferred mobile ion concentrations increase by more than 100$\times$, to over $1\times10^{18}$ cm$^{-3}$, within just three minutes of reverse bias at $-6.0$ V in p-i-n perovskite diodes. We attribute this increase to iodide oxidation and coupled iodine vacancy creation that must be balanced by reduction reactions near the HTL. Sub-optimal HTL coverage leads to direct contact between the transparent conducting electrode and perovskite and facilitates reduction, enables the creation of even larger inferred mobile ion concentrations ($\sim1\times10^{19}$ cm$^{-3}$), and leads to faster degradation under reverse bias. This explains previous work that showed increased breakdown voltages and improved reverse-bias stability by implementing thick, uniform HTLs. 

\section*{KEYWORDS}


perovskite, diode, solar cells, reverse bias, tunneling, electrochemistry, ion migration

\section*{INTRODUCTION}

Perovskite solar cells (PSCs) have achieved 27\% power conversion efficiencies (PCEs), competing with record silicon single-junction solar cells \cite{NREL_efficiency_chart,chen_improved_2024}. Their stability under lab-controlled light plus heat testing is rapidly improving \cite{jamesh_advancement_2025}, but several instability challenges still exist related to chemical migration, photochemistry, and electrochemistry between perovskite elements and the surrounding materials \cite{zhao_redox_2016, domanski_not_2016, li_direct_2017, kim_large_2018, kerner_electrochemical_2019, kerner_low_2020, bertoluzzi_incorporating_2021, ni_evolution_2021, xu_halogen_2023, xu_origins_2023, xu_beyond_2024, zhan_indium_2024}. These challenges must be fully addressed for PSCs to exhibit the desired 25-year outdoor operational lifetimes. As in other thin-film photovoltaic technologies, PSCs show concerning instabilities under \emph{reverse bias}, a condition that can be experienced by shaded cells in a series-connected module \cite{seyedmahmoudian_simulation_2015, qian_destructive_2020, bertoluzzi_incorporating_2021, xu_reverse-bias_2023, islam_modeling_2025}. PSCs often abruptly pass large localized current densities and become permanently shorted at small reverse potentials of just $-1$ V to $-4$ V \cite{bowring_reverse_2018, razera_instability_2019, li_sparkling_2022}. This type of breakdown has been alleviated by using less reactive electrodes \cite{bogachuk_perovskite_2022, jiang_improved_2024, lanzetta_tinlead_2025}, by minimizing the density of pinholes in the perovskite \cite{johnson_how_2025}, and by using thick, dense transport layers that do not have thin, electrically weak spots where breakdown can occur \cite{tan_stability_2020, li_barrier_2024, jiang_improved_2024, amorales_nickel-oxide_2026}. When cells do \emph{not} abruptly break down, they instead show a \emph{gradual} increase in reverse current with increasing reverse potential and time spent under reverse bias. Still, the reverse current drives electrochemical reactions, including the oxidation of iodine and metallic electrodes, leading to fast degradation of the solar cell \cite{bertoluzzi_incorporating_2021, ni_evolution_2021, bogachuk_perovskite_2022, henzel_impact_2023, li_barrier_2024, ren_mobile_2024, lanzetta_tinlead_2025}. Iodine complexes resulting from iodide oxidation under reverse bias have been shown to leave the perovskite \cite{ren_mobile_2024} or the entire solar cell \cite{bogachuk_perovskite_2022}. Oxidation of halide species is thus a central challenge facing reverse-bias stability in perovskite diodes \cite{razera_instability_2019, bertoluzzi_incorporating_2021, bogachuk_perovskite_2022, xu_halogen_2023, henzel_impact_2023, li_barrier_2024, ren_mobile_2024, jiang_improved_2024}. These previous results highlight that perovskite solar cells must be considered as both electrochemical cells and semiconductor devices. They have limited voltage ranges over which they are stable, and operation outside of this stability window leads to rapid degradation. \\

A popular model for explaining gradual reverse breakdown in perovskite diodes describes mobile ions drifting under the high internal electric fields to create space charge and steep band bending that enables electronic charge tunneling \cite{bertoluzzi_mobile_2020, schmidt_characterization_2025, bao_reversebias_2025}. Iodide ions (I$^-$) are the dominant mobile ion species, and their transport is mediated by iodine vacancies \cite{eames_ionic_2015, senocrate_nature_2017, kim_large_2018, weber_how_2018, birkhold_direct_2018, bertoluzzi_mobile_2020}. For convenience, iodine vacancies can be treated as mobile \emph{cations}. In a typical p-i-n architecture, a transparent conducting oxide (TCO) is covered by a hole-transport layer (HTL), and the back contact includes an electron-transport layer (ETL) capped with a metal electrode. Under reverse bias, iodine vacancies accumulate in a very thin layer (Debye layer) near the HTL, leading to a negatively charged vacancy \emph{depletion} region, strong band bending, and shorter charge-tunneling barriers near the ETL \cite{bertoluzzi_mobile_2020, schmidt_characterization_2025, bao_reversebias_2025}. Reverse current then arises from hole tunneling from the back metal electrode into the perovskite valence band \cite{bowring_reverse_2018, bertoluzzi_mobile_2020, bao_reversebias_2025}. However, previous experiments indicate typical concentrations of ionic defects \cite{de_keersmaecker_defect_2024} and mobile ions of about $10^{17}$ cm$^{-3}$ \cite{almora_capacitive_2015, bertoluzzi_situ_2018, bertoluzzi_mobile_2020, penukula_quantifying_2023} or smaller \cite{weber_how_2018, birkhold_direct_2018, penukula_quantifying_2023}. (Supplemental Information [SI] Section 2 summarizes previously reported mobile ion concentrations). According to calculations shown in this report, mobile ion concentrations of $10^{17}$ cm$^{-3}$ or smaller lead to hole tunneling barrier widths of about 35 nm or larger at $-5$ V, too wide to quantitatively explain experimentally observed reverse current densities. Additionally, the hole-tunneling model in its present form does not easily explain recent experimental observations that the \emph{HTL} significantly impacts the reverse-bias behavior \cite{jiang_improved_2024, zhao_ambient_2025}. Thick polymeric HTLs substantially increase reverse-breakdown voltages ($V_\text{br}$) \cite{jiang_improved_2024, zhao_ambient_2025}, reaching about $-15$ V when 35 nm of poly[bis(4-phenyl)(2,4,6-trimethylphenyl)amine (PTAA) was used \cite{jiang_improved_2024}. Excellent HTL uniformity also leads to larger $V_\text{br}$ and more gradual breakdown behavior \cite{zhao_ambient_2025, amorales_nickel-oxide_2026}. The connection between the HTL design and hole tunneling through the opposite transport layer likely lies in the law of charge conservation in electrochemical processes \cite{jiang_improved_2024}. The perovskite can be thought of as a solid-state electrolyte, especially if mobile ions are concentrated in grain boundary and interfacial regions where local concentrations can be as high as 1 mM, and oxidation must be balanced by reduction elsewhere in the cell \cite{jiang_improved_2024}. Balanced oxidation-reduction (redox) reactions at opposite electrodes have also been proposed to explain phase segregation in mixed-halide perovskites under bias \cite{xu_halogen_2023}. In this regard, a thick HTL that eliminates perovskite contact with the TCO should be a very good electron-blocking layer; it will \emph{directly} slow reduction reactions and thus \emph{indirectly} slow the paired oxidation reactions under reverse bias. The oxidation of iodide generates interstitial iodine and, important for this study, \emph{new iodine vacancies}, both of which strongly influence the cell's optoelectronic properties and degradation \cite{kim_large_2018,motti_controlling_2019,ni_evolution_2021,diekmann_determination_2023,thiesbrummel_ion-induced_2024,seid_mitigating_2025}. However, the impact that iodide oxidation and newly created mobile vacancies have on reverse-bias breakdown behaviors and stability have so far not been established. \\

For this study, we hypothesized that the previous hole-tunneling model remains valid but requires a critical modification: the mobile ion concentration is not fixed and instead evolves as iodide oxidation proceeds. Since oxidation and reduction are paired, the HTL indirectly regulates oxidation rates by directly regulating reduction rates at the TCO-perovskite interface. We validate this hypothesis by measuring mobile ion concentrations before and immediately after fixed reverse bias stress using ionic current-transient measurements \cite{bertoluzzi_situ_2018, bertoluzzi_mobile_2020, thiesbrummel_ion-induced_2024, schmidt_characterization_2025}. In our devices, inferred mobile ion concentrations increase by over 100$\times$, from $3\times$10$^{15}$ cm$^{-3}$ to $1.2\times$10$^{18}$ cm$^{-3}$, under reverse bias at $-6$ V within just three minutes. These high mobile ion concentrations \emph{dominate} the band bending, resulting in high reverse electron/hole tunneling current densities seen in experiments. We calculate current density vs. voltage ($J-V$) curves using a simplified charge-tunneling model and find good agreement with measurements only when the mobile ion concentration is allowed to increase rapidly with time spent under reverse bias. We then discuss and experimentally demonstrate \emph{why} the HTL impacts reverse-bias behavior. If the HTL blocks electron transfer from the TCO electrode to the perovskite, it will slow reduction reactions and limit the generation of mobile ions, thereby maintaining high $V_\text{br}$ and improving reverse-bias stability. We show that typical PTAA HTLs do not completely cover tin-doped indium oxide (ITO) electrodes because of ITO's nanometer-scale surface texture. Incomplete HTL coverage facilitates reduction reactions under reverse bias, leading to up to $10\times$-larger inferred mobile ion concentrations ($1\times$10$^{19}$ cm$^{-3}$) than smooth HTL-coated substrates, small $V_\text{br}$, and rapid cell degradation under fixed reverse bias. \\

\section*{RESULTS}
\begin{figure}[h!]
      \includegraphics[width=5.5in]{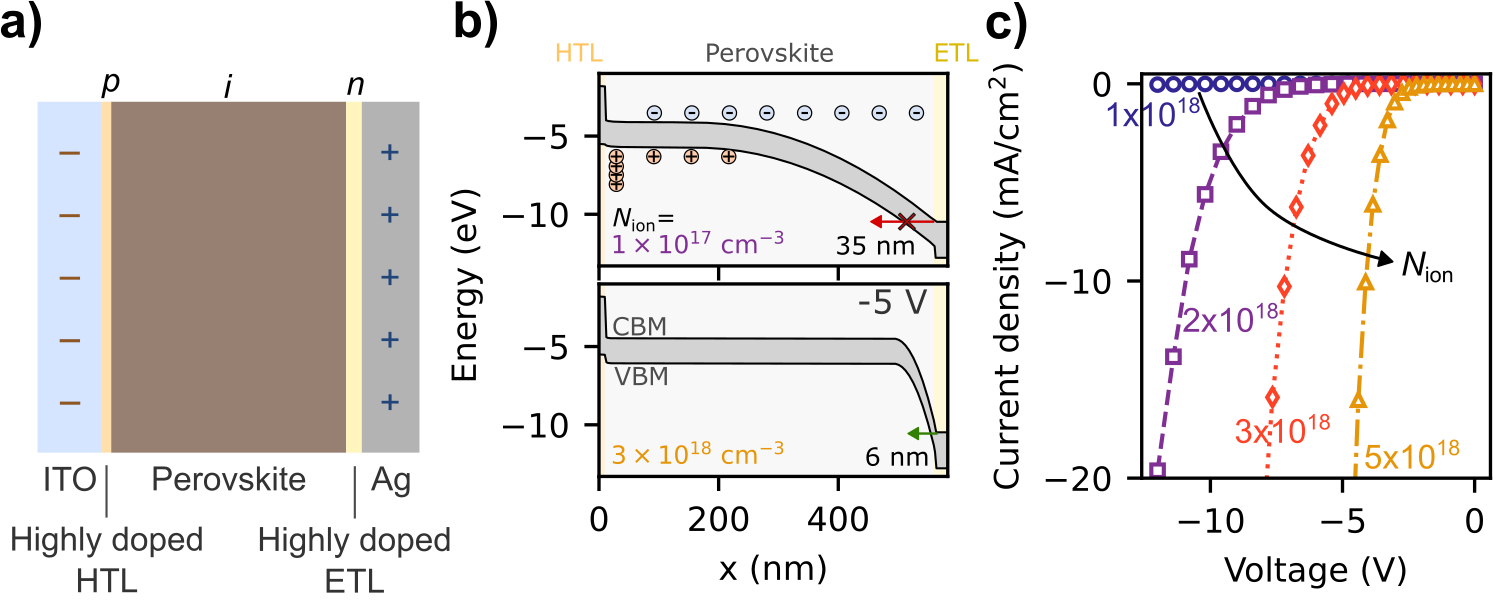}
	\caption{
        (\textbf{a}) Idealized p-i-n perovskite diode architecture used to illustrate the relationship between mobile ion concentration and reverse breakdown voltages. `+' and `-' symbols in electrodes indicate the relative potentials applied under reverse bias.
        (\textbf{b}) Calculated band diagrams for cells held at a reverse bias of $-5$ V, assuming a single mobile ion species (cations, representing iodine vacancies) and highly doped transport layers. Schematic ion distributions (circles with `+' and `-' symbols) are shown. Calculations for two different mobile cation concentrations $N_\text{ion}$ are shown. Top: $N_\text{ion}=1\times10^{17}$ cm$^{-3}$. Bottom: $N_\text{ion}=3\times10^{18}$ cm$^{-3}$. Hole tunnel barrier widths are specified in each panel. CBM = conduction-band minimum. VBM = valence-band maximum.
        (\textbf{c}) Calculated reverse $J-V$ curves from a charge-tunneling model at the ETL for band diagrams like those shown in (b) with mobile ion concentrations $N_\text{ion}$ varying between $1\times10^{18}$ cm$^{-3}$ and $5\times10^{18}$ cm$^{-3}$.
        }
\label{Fig:Calculations}
\end{figure}

To first develop intuition for the relationship between mobile ion concentrations and reverse $J-V$ behavior, we calculate band diagrams and $J-V$ curves for the idealized device stack shown in Fig. \ref{Fig:Calculations}a based on the band-bending model described in the introduction \cite{bertoluzzi_mobile_2020}. The model includes mobile cations (representing highly mobile iodine vacancies), fixed (or, at least, much less mobile) counter-ions (anions), and metallic contact layers \cite{bertoluzzi_mobile_2020, calado_driftfusion_2022}. Fig. \ref{Fig:Calculations}b shows band diagrams at $-5$ V, highlighting the ionic depletion and band bending near the ETL. (See SI Section 3 for more information on band-diagram calculations.) A mobile ion concentration $N_\text{ion}=1\times10^{17}$ cm$^{-3}$ (top panel) is comparable to previously reported concentrations \cite{almora_capacitive_2015, bertoluzzi_situ_2018, bertoluzzi_mobile_2020, penukula_quantifying_2023} and gives a tunneling barrier width of about 35 nm at $-5$ V in this model. Fig. \ref{Fig:Calculations}c shows calculated $J-V$ curves using a Schottky charge-tunneling model \cite{chang_carrier_1970} and band diagrams calculated using the ionic depletion approximation \cite{bertoluzzi_mobile_2020} for a range of mobile ion concentrations. (See SI Section 4 for tunneling-current calculation details.) These calculations show that a 35-nm tunneling barrier is too wide to enable appreciable tunneling current densities. In other words, gradual breakdown with $V_\text{br} \approx -5$ V commonly observed in perovskite p-i-n diodes is not easily explained by band bending due to mobile ions at a constant concentration of approximately $10^{17}$ cm$^{-3}$ \cite{bowring_reverse_2018, bertoluzzi_mobile_2020, schmidt_impact_2023, schmidt_characterization_2025}. When $N_\text{ion}$ increases to $3\times10^{18}$ cm$^{-3}$, the tunneling barrier \emph{in this model} decreases to about 6 nm (Fig. \ref{Fig:Calculations}b bottom panel) and calculated $J-V$ curves (Fig. \ref{Fig:Calculations}c, red diamonds) become comparable to experimentally observed $J-V$ curves with $V_\text{br}$ of about $-5$ V. Importantly, Fig. \ref{Fig:Calculations}c indicates that $V_\text{br}$ continuously decreases as the mobile cation concentration increases. In this model, mobile ion concentrations should be about $3\times10^{18}$ cm$^{-3}$ to explain $V_\text{br}$ values of about $-5$ V. SI Section 5 describes how oxidizing just a single layer in a 500-nm thick perovskite film would result in nearly $3\times$10$^{18}$ cm$^{-3}$ volume-average ion concentration (Fig. S1). This mobile ion concentration increase is also reasonable based on previous reports that showed increased ionic conductivities under illumination, increased electronic trap densities under reverse bias, and performance degradation under standard operating conditions \cite{kim_large_2018, ni_evolution_2021, thiesbrummel_ion-induced_2024, seid_mitigating_2025}. \\

\begin{figure}[h!]
    \includegraphics[width=6.6in]{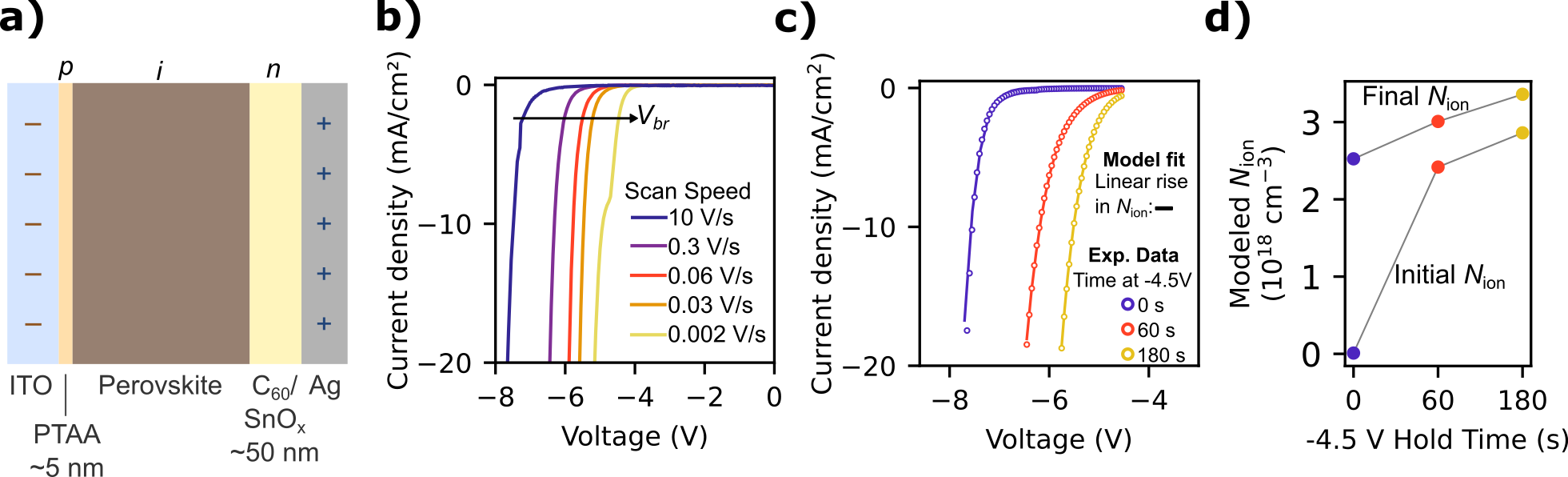}
	\caption{
        (\textbf{a}) P-i-n perovskite diode architecture used in our main measurements in this work.
        (\textbf{b}) Experimental reverse $J-V$ scans with varying scan rates between 10 V/s and 0.002 V/s.
        (\textbf{c}) Experimental reverse $J-V$ scans (circle markers) measured with a scan rate of 4 V/s after holding $-4.5$ V across the cell for 0 s (blue), 60 s (red), and 180 s (gold). Experimental curves are fit using a hole tunneling model assuming a linearly increasing $N_\text{ion}$ (solid curves).
        (\textbf{d}) Summary of inferred mobile ion concentrations from tunneling-current model fits to experimental curves shown in panel (b) after holding cells at $-4.5$ V for 0 s, 60 s, and 180 s. Concentrations at the start of the scan (``initial") and at the end of the scan (``final") are shown.
        }
\label{Fig:Observations}
\end{figure}

Our hypothesis that mobile ion concentrations increase under reverse bias is initially explored by performing a series of reverse $J-V$ scans of PSCs with scan rates ranging from 10 V/s to 0.002 V/s (spanning nearly four orders of magnitude) in the dark. All results presented in this manuscript were performed on cells with the following architecture: ITO/PTAA/Cs$_{0.25}$FA$_{0.75}$Pb (I$_{0.85}$Br$_{0.15}$)$_3$/C$_{60}$/SnO$_x$/Ag (except where otherwise specified; Methods), illustrated in Fig. \ref{Fig:Observations}a. A 30-nm-thick dense SnO$_x$ layer was used to mitigate electrochemistry between the perovskite and metal contact \cite{li_barrier_2024}. Experiments were performed on 200-$\mu$m-diameter circular cells to obtain consistent gradual reverse-bias behavior by avoiding defect-mediated breakdown \cite{johnson_how_2025}. Fig. \ref{Fig:Observations}b shows experimental $J-V$ curves recorded with varying scan rates. The fastest scan rate ($10$ V/s) requires less than one second to complete, while the slowest scan rate requires over 30 minutes. Importantly, each measurement was performed on a fresh cell. The $V_\text{br}$ (defined here as the voltage at which the current density reaches $-2$ mA/cm$^2$) continuously decreases from about $-8$ V to $-5$ V as the scan rate decreases (\emph{i.e.} $V_\text{br}$ moves towards $0$ V). This effect was repeatable across different batches and cell architectures (Fig. S2). A similar effect was observed by performing multiple reverse $J-V$ scans at a constant scan rate on the same cell (Fig. S3). Previous reports \cite{bao_reversebias_2025} and drift-diffusion model calculations (Fig. S4) indicate that scan rates below $10$ V/s begin to probe the system with mobile iodide and iodine vacancies in electrostatic equilibrium with the applied field at each voltage step of the measurement \cite{bertoluzzi_situ_2018, calado_driftfusion_2022}. Therefore, the observed monotonic decrease in $V_\text{br}$ for scan speeds down to 0.002 V/s likely arises from a \emph{changing mobile ion concentration} rather than slow ion dynamics. As standards for measuring and reporting $V_\text{br}$ values in perovskite cells are developed, it should be noted that $V_\text{br}$ depends on the scan rate. \\

During the scans shown in Fig. \ref{Fig:Observations}b, both the applied potential and elapsed time continuously increase. To better distinguish the effects of potential and duration, we first held devices at a \emph{fixed} reverse potential of $-4.5$ V for 0 s, 60 s, and 180 s and then immediately performed fast reverse $J-V$ scans. We chose a fast scan rate of approximately 4 V/s (requiring about one second to complete the scan) to minimize the increase in the mobile ion concentration during the scan. Additionally, since iodide should completely drift into equilibrium within about one second \cite{schmidt_characterization_2025}, our hold times of 60 seconds or longer are sufficiently long to avoid changes solely due to mobile iodide drift. Changes in $J-V$ curves across our measurements are thus expected to be dominated by changes in the mobile ion concentration. Fig. \ref{Fig:Observations}c shows experimental reverse $J-V$ curves for initial hold times of 0 s (blue markers), 60 s (red markers), and 180 s (gold markers). As the hold time increases, $V_\text{br}$ decreases from about $-7$ V to $-5$ V, confirming that longer \emph{time under reverse bias} causes the decrease in $V_\text{br}$. To better understand the experimental trend, we calculated tunneling-current $J-V$ curves to best represent each experimental curve. First, we assumed a \emph{fixed} $N_\text{ion}$ for each curve (Fig. S5) but the fit quality was poor, likely due to the mobile ion concentration changing substantially during the experimental scan. To account for that effect, we calculated tunneling-current $J-V$ curves allowing $N_\text{ion}$ to \emph{increase} with reverse bias for each curve (Fig \ref{Fig:Observations}c solid curves). Although the precise dependence of the increase in $N_\text{ion}$ on applied potential and time is not \emph{a priori} known, a simple linear $N_\text{ion}$ increase improves the fit quality substantially for all hold times. This is reasonable given the very fast $J-V$ scan that might sample small, nearly linear behavior of more complicated functional behavior. The resulting fit $N_\text{ion}$ values indicate very large and fast increases in mobile ion concentrations. Fig. \ref{Fig:Observations}d shows the initial (at the start of the scan) and final (at the end of the scan) $N_\text{ion}$ estimates. As the hold time increases, the \emph{initial} $N_\text{ion}$ value increases from $1\times10^{16}$ cm$^{-3}$ to $2.9\times10^{18}$ cm$^{-3}$ within just 60 seconds of fixed reverse bias at $-4.5$ V and additional time does not lead to significant further increases in $N_\text{ion}$. That is, $N_\text{ion}$ quickly reaches a saturation level of about $3\times10^{18}$ cm$^{-3}$. Calculated tunneling currents are very sensitive to variations in energy level alignments and perovskite electric permittivity (SI Section 4) \cite{aeberhard_multi-scale_2024, bao_reversebias_2025}. We analyzed the $J-V$ curves shown in Fig. \ref{Fig:Observations}c using the Schottky tunneling model described in Fig. \ref{Fig:Calculations}, and we therefore want to emphasize the \emph{trends} seen in the inferred $N_\text{ion}$ values that are expected to be more robust than the absolute $N_\text{ion}$ values \cite{bao_reversebias_2025}. \\

\begin{figure}[h!]
      \includegraphics[width=4.5in]{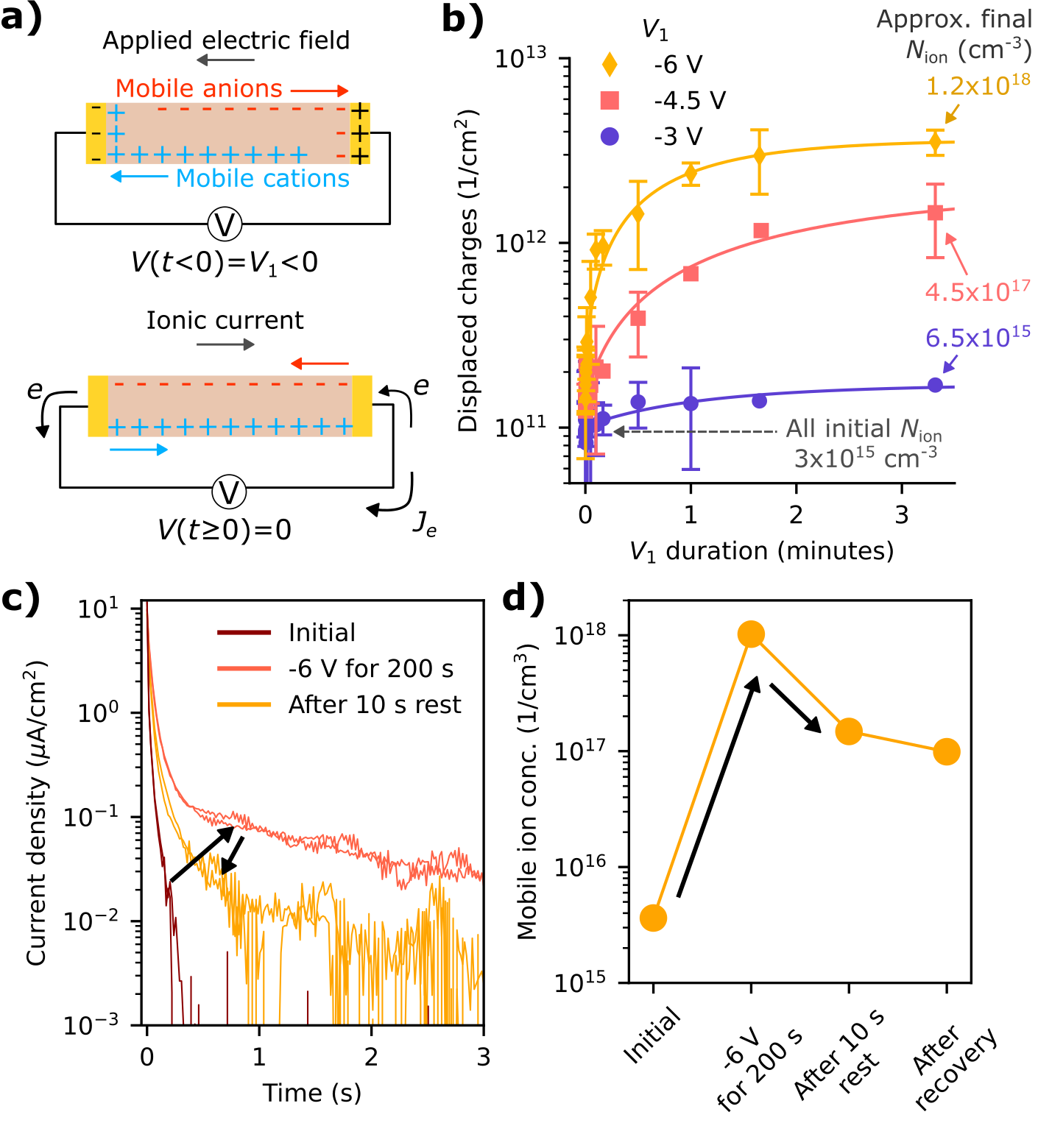}
	\caption{
        (\textbf{a}) Schematic of ionic current-transient measurements. Devices are held at $V_1<0$ (top panel) and then quickly switched to $V_2=0$ (bottom panel). Ionic motion in the cell after switching to 0 V leads to a measurable external transient current. 
        (\textbf{b}) Measured external integrated current (displaced charges) in a PSC after holding $V_1=-3.0$ V (blue), $V_1=-4.5$ V (red), and $V_1=-6.0$ V (gold) for varying durations between about 100 ms and 3.5 minutes. Inferred initial and final mobile ion concentrations are indicated by text near each data set. Markers are measured values. Error bars represent $3\sigma$ spread in measured values. Empirical model fits to a single exponential saturating function $Q(t_1)\propto1-\exp[-t/\tau]$ are shown as solid curves. 
        (\textbf{c}) Measured current densities during the current transient immediately after holding the cell at $V_1=-6.0$ V for 300 ms (brown), for 200 s (red), and then again for 300 ms (gold). 
        (\textbf{d}) Evolution of mobile ion concentrations immediately after holding at $-6$ V for 200 s, after resting for about 10 seconds in the dark, and after a ``recovery" process. Black arrows in (c) and (d) indicate the measurement progression.
        }
\label{Fig_Mobile_Ion_Measurements}
\end{figure}

To more directly measure mobile ion concentrations, we implemented an ionic current-transient measurement that has been previously used with perovskites \cite{bertoluzzi_situ_2018, bertoluzzi_mobile_2020, penukula_quantifying_2023, thiesbrummel_ion-induced_2024, schmidt_how_2025, schmidt_characterization_2025}, modified here to estimate mobile ion concentrations immediately after reverse bias. We first hold cells under a reverse bias ($-3$ V to $-6$ V), during which mobile ions drift across the perovskite under the applied electric field and equilibrate to screen the internal field (Fig. \ref{Fig_Mobile_Ion_Measurements}a, top) \cite{bertoluzzi_incorporating_2021}. The applied potential is then quickly switched (over milliseconds) to 0 V. Mobile ions drift and equilibrate to this new potential, producing an internal ionic displacement current and a measurable external electronic current (Fig. \ref{Fig_Mobile_Ion_Measurements}a, bottom). The current measured during the first $\approx$20 ms at 0 V is excluded to avoid current from electronic carriers that are expected to be largely extracted within the first few microseconds to milliseconds \cite{bertoluzzi_mobile_2020}. The transient current after that duration, and extending out to several seconds, arises predominantly from internal ionic drift (rather than electric contributions) because the applied potential is 0 V \cite{bertoluzzi_situ_2018, bertoluzzi_mobile_2020, thiesbrummel_ion-induced_2024}. Integrating the current transient over time directly gives the total \emph{external displaced electronic charge} during ion equilibration. The experimental method is based on the fact that the electrode charge needed for the solar cell to be at a certain voltage depends on its capacitance, which varies as a function of the mobile ion concentration because mobile ions can completely screen the internal field. When the mobile ion concentration is above about $10^{18}$ cm$^{-3}$ to $10^{19}$ cm$^{-3}$, the ionic depletion and accumulation regions are less than a few nanometers (much less than the device thickness) and the capacitance varies only slightly with the mobile ion concentration \cite{schmidt_how_2025}. For this reason, these methods can accurately measure the mobile ion concentration only if it is less than a few-times-$10^{18}$ cm$^{-3}$ and yield increasingly large uncertainties for higher concentrations. We built an analytical model using the one-mobile-ion distribution proposed by Bertoluzzi \emph{et al.} \cite{bertoluzzi_mobile_2020} to relate the total external displaced electronic charge to the internal \emph{mobile ion concentration} (SI Section 6; Fig. S9). This model accounts for the finite thickness and relatively low dielectric constants of the transport layers, which leads to a notable portion of the potential being dropped across the ETL. (Little potential is dropped across the HTL since it is very thin in this work.) We hold various reverse potentials for durations ranging from about 100 ms to 200 s (over three minutes) to directly assess the evolution of mobile ion concentrations in PSCs \emph{immediately} after reverse bias. \\

To first establish that our system reliably estimates mobile ion concentrations comparable with previous literature, we first perform the ``standard" measurement with $V_1=+0.8$ V (before reverse-bias stress), from which we infer mobile ion concentrations to be between about $3\times10^{16}$ cm$^{-3}$ and $1\times10^{17}$ cm$^{-3}$ (Fig. S6)\cite{bertoluzzi_situ_2018, bertoluzzi_mobile_2020, penukula_quantifying_2023, thiesbrummel_ion-induced_2024}. The inferred mobile ion concentration does however depend on the $V_1$ hold time, decreasing to about $3\times10^{15}$ cm$^{-3}$ in our cells for longer hold times. We do not expect the forward current densities of approximately 10 to 100 $\mu$A/cm$^2$ at $+0.8$ V to be harsh enough to cause device degradation \cite{thiesbrummel_ion-induced_2024}. The device may instead be recovering from intrinsic mobile defects under these conditions due to the injection of both electrons and holes \cite{motti_controlling_2019}. \\

Fig. \ref{Fig_Mobile_Ion_Measurements}b summarizes the evolution of \emph{current-transient measurements} after fixed \emph{reverse bias}. We directly plot the measured external displaced charge and provide initial and final $N_\text{ion}$ estimates by text near each curve assuming the one-mobile-cation model described above and in SI Section 6. With $V_1=-3.0$ V, which is smaller than $V_\text{br}$ (Fig. S7), $N_\text{ion}$ increases only very slightly from about $3\times10^{15}$ cm$^{-3}$ to $6.5\times$10$^{15}$ cm$^{-3}$ over three minutes (blue markers). We tested two larger reverse potentials. We chose $V_1=-4.5$ V (red markers) to be at the ``onset" of reverse breakdown with typical current densities of $\sim$1 $\mu$A/cm$^2$ (Fig. S7). We chose $V_1=-6.0$ V (gold markers) because it provides even larger reverse currents of several 100 $\mu$A/cm$^2$ to 1 mA/cm$^2$ (Fig. S8). With $V_1=-4.5$ V, $N_\text{ion}$ starts to increase sharply around 10 s and reaches a final value of about $4.5\times10^{17}$ cm$^{-3}$. With $V_1=-6.0$ V, $N_\text{ion}$ starts increasing earlier, near three seconds, and reaches a higher maximum value of $1.2\times10^{18}$ cm$^{-3}$. (Fig. S8 presents this data on a log-log scale to better highlight timescales of the ion creation process.) Importantly, the timescales for the large inferred $N_\text{ion}$ increase (3 to 100 s) are much longer than the durations that would be needed for mobile iodide and iodine vacancies to equilibrate (about 100-300 ms). Although the inferred mobile ion concentration depend on the assumed internal distribution of mobile ions, the permittivity of the perovskite, and the thickness and permittivities of the transport layers \cite{bertoluzzi_mobile_2020} (Fig. S9), all models indicate an approximately $400\times$ \emph{increase} in mobile ion concentrations after holding at $-6$ V for three to five minutes.  \\

A few general results from our mobile ion concentration measurements are worth emphasizing. First, $N_\text{ion}$ saturates after several hundred seconds in all cases, similar to the behavior and timescales inferred from analyses of reverse $J-V$ curves shown in Figs. \ref{Fig:Observations}c,d. Second, the rate of mobile ion \emph{creation} increases with larger reverse-bias voltages. The gradual reverse breakdown at about $-8$ V even for the fastest scan rates seen in Fig. \ref{Fig:Observations}b can therefore happen because mobile ions are generated very quickly once the reverse bias is high enough to drive significant reverse current densities (greater than $\sim$100 $\mu$A/cm$^2$). \\

A closer inspection of the current densities over the transient period ($t>0$) provides useful insight into the dynamics of the mobile ions contributing to the increased mobile ion current. Fig. \ref{Fig_Mobile_Ion_Measurements}c shows the transient current densities before and after reverse-bias stressing. The `initial' curve is the current transient measured after holding the cell at $-6.0$ V for 300 ms (brown) for the first time. The current density decays almost completely within 500 ms with a characteristic (exponential) time constant of about 35 ms (Fig. S10). The current transient immediately after holding the cell at $-6.0$ V for 200 s (red) shows over $100\times$ greater current density over most of the decay, with possible indications of a second, slower decay component with a characteristic time constant of about 2.3 seconds that becomes the dominant component after about 500 ms (Fig. S10). Although the current density after 500 ms is small ($\sim100$ nA) compared to the peak currents ($\sim$20 $\mu$A), it contributes substantially to integrated current because it persists for about 10 seconds. By comparison with drift-diffusion calculations previously published, this long decay component appears to persist much longer than expected for mobile ions with a mobility of about $10^{-9}$ cm$^2$/Vs \cite{schmidt_characterization_2025, schmidt_how_2025}; this decay behavior will be discussed in greater detail later when we hypothesize specific redox processes occurring during reverse bias. A final measurement was performed after letting the same cell rest in the dark for about 10 seconds (`After 10 s rest', gold). The decay approaches the initial behavior but with about $10\times$ increased current density over most of the decay. This new contribution appears to be from permanent, or at least long-living, new mobile ions \cite{de_keersmaecker_activated_2025}. To better understand this hysteresis, we attempted to recover the mobile ion concentration back to starting values by putting the cell in moderate forward-bias conditions in the dark, over which the cell passed $\sim$ mA/cm$^2$ current densities until the forward $J-V$ behavior stopped evolving. A final mobile ion concentration of about $1\times10^{17}$ cm$^{-3}$ was inferred after this ``recovery" process, still about 25 times larger than the initial value. Fig. \ref{Fig_Mobile_Ion_Measurements}d summarizes this evolution in mobile ion concentrations. Since the data indicates a large concentration of newly oxidized/reduced species that subsequently relax over less than 10 seconds, a portion of the transient current at 0 V (Fig. \ref{Fig_Mobile_Ion_Measurements}c) may be faradaic current from oxidation (reduction) near the HTL (ETL). The current transients more closely follow double-exponential-like decay behavior consistent with a two-mobile-ion reorganization (a capacitive effect) rather than $\propto t^{-1/2}$ behavior of diffusion-limited redox current (Fig. S10) \cite{de_keersmaecker_how_2022}. Both interpretations, however, indicate an increase in mobile ion concentrations. Alternatively, paired mobile ions in the cell could annihilate without causing an external current. \\

\begin{figure}[h!]
      \includegraphics[width=6.5in]{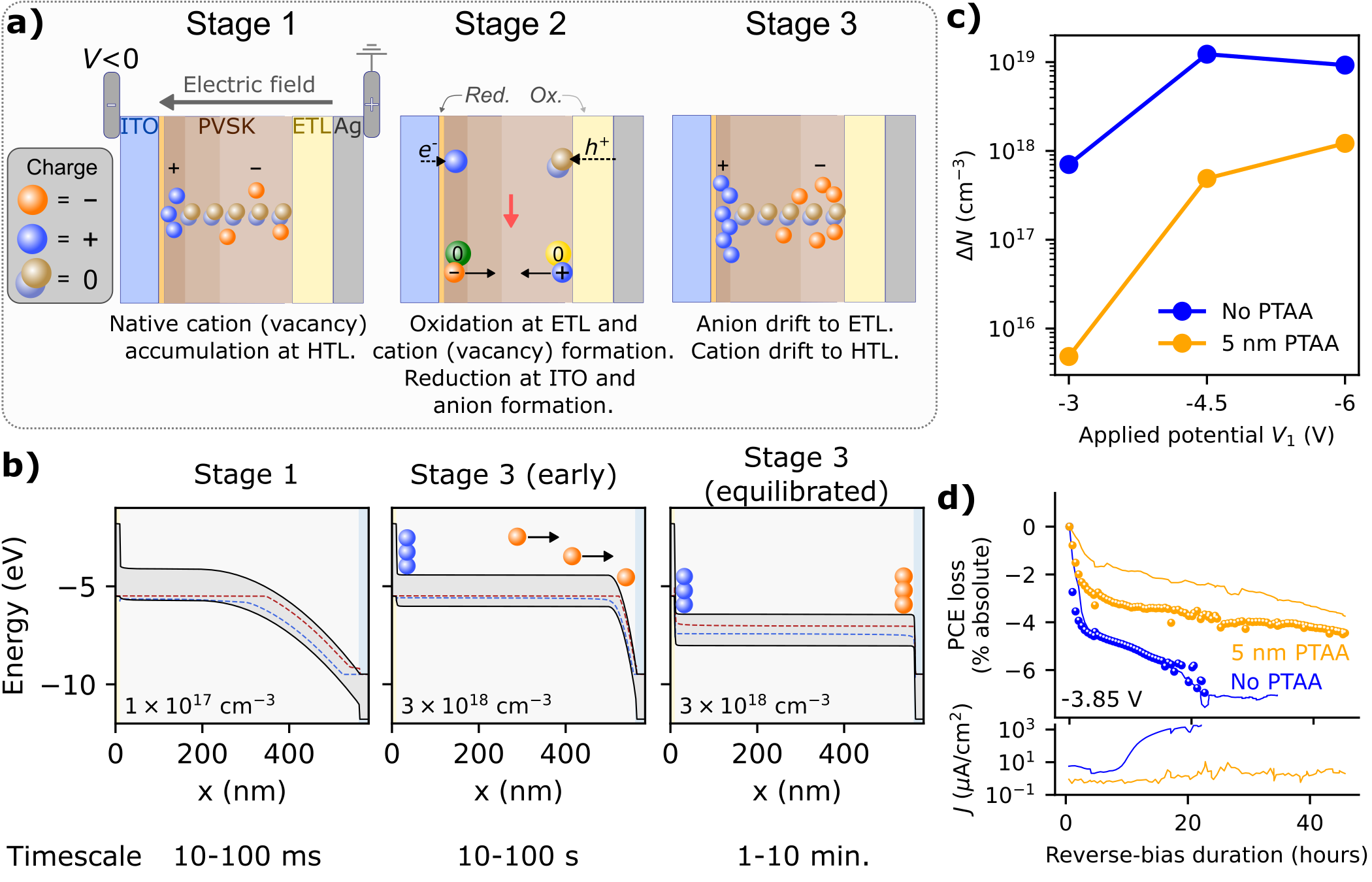}
	\caption{
        (\textbf{a}) Illustration of balanced reduction-oxidation processes in perovskite diodes under reverse bias. The process is divided into three `Stages'. Blue circles = positive ions (V$_\text{I}^+$, equivalent to under-coordinated Pb$^{+x}$).  Orange circles = negative ions (I$^-$ in Stage 1, I$^-$ and V$_\text{Pb}^-$ in Stages 2 and 3). e$^-$ = electron. h$^+$ = hole. Clustered blue plus brown-orange circles represent a stoichiometric unit cell.
        (\textbf{b}) Energy band diagrams showing the perovskite conduction and valence bands (solid black curves), along with electron and hole Fermi levels (blue and red dashed curves, respectively) for Stage 1 (left), Stage 3 before electrostatic ion equilibration (center), and Stage 3 after ion equilibration (right). New mobile ions (orange and blue circles) are schematically shown. Tentative timescales inferred from experiments are presented below each illustrated process. 
        \textbf{c}) Inferred mobile ion concentrations immediately after 3.5 minutes of fixed reverse bias at three different reverse biases. Results from PSCs with a 5-nm-thick PTAA HTL (orange) and without an HTL (blue) are shown.
        \textbf{d}) Top panel: PCE decay (absolute \%) of PSCs with a 5 nm PTAA HTL (orange) and no HTL (blue) under fixed reverse bias at $-3.85$ V. Bottom panel: Reverse current densities flowing through each cell during the fixed voltage holds. 
        }
\label{FigModel}
\end{figure}

To consolidate our experimental observations, we propose a three-stage electrochemistry-based model illustrated in Fig. \ref{FigModel}a. Under reverse bias, the ITO is held at a negative potential ($V<0$) compared to the top metal electrode. `\emph{Stage 1}' shows the initial ion configuration. The perovskite has a ``native" mobile iodine vacancy concentration (V$_I^+$) of about $1\times 10^{17}$ cm$^{-3}$ \cite{bertoluzzi_mobile_2020, de_keersmaecker_defect_2024}. The V$_I^+$ will quickly accumulate at the HTL interface under reverse bias. This equivalently corresponds to a high concentration of under-coordinated Pb (Pb$^{+x}$; here represented by blue circles) at the HTL and an excess of iodide (I$^-$) near the ETL (orange circles in \emph{`Stage 1'}). This stage of the model is identical to the model proposed by Bertoluzzi \emph{et al.} \cite{bertoluzzi_mobile_2020}. In `\emph{Stage 2}', iodide is oxidized. Iodide oxidation is not necessarily confined to the ETL interface; like iodide oxidation under illumination \cite{kerner_role_2021, xu_correlating_2024}, the oxidation event under reverse bias might happen in the bulk due to small hole-tunneling currents and an excess of holes from \emph{Stage 1} \cite{bertoluzzi_incorporating_2021}. In either case, iodide oxidation generates iodine (yellow circle, possibly I$_2$, or ultimately I$_3^-$) and, critically, a new positively charged iodine vacancy (V$_I^+$) as the iodine dissociates from the lattice \cite{kim_large_2018, motti_controlling_2019, bogachuk_perovskite_2022}. The neutral iodine will diffuse through the perovskite and either become trapped in the HTL \cite{kerner_organic_2021} or the ETL \cite{kim_large_2018, ni_evolution_2021, ren_mobile_2024, razera_instability_2019, xu_halogen_2023}, or diffuse through the ETL and corrode the top metal contact if a dense, impermeable metal oxide ETL layer is not used\cite{zhao_redox_2016, li_direct_2017, bogachuk_perovskite_2022, xu_origins_2023} (unlike in cells with ALD SnO$_x$). Neutral iodine does not directly impact band bending, and so its final location will be ignored for now. Critically, a reduction event must simultaneously occur somewhere in the cell to maintain charge balance. Very few electrons are available in the bulk of the perovskite under reverse bias \cite{bertoluzzi_incorporating_2021}. Reduction is therefore likely to be confined to the negative electrode (TCO) interface due to the very high concentration of electrons there and is enabled by ITO-perovskite contact due to protrusions on the ITO surface and pinholes in the HTL as we demonstrate later. For band bending to \emph{increase} near the ETL during this paired reduction-oxidation process (consistent with experimentally observed decreased $V_\text{br}$), a new mobile \emph{anion} must arrive at the ETL (orange circles represent I$^-$ \emph{and} the new anion species at this stage). Therefore, we speculate that the reduction event generates a new mobile (but possibly very slow) \emph{anion} species. In `\emph{Stage 3}', new (highly mobile) iodine vacancies quickly accumulate at the HTL interface. The new mobile anion species will slowly drift toward the ETL and the top metal contact. \\

The specific band diagram evolution resulting from the paired reduction-oxidation processes depends on whether the new reduced species is mobile or fixed, what iodine complexes are forming, and whether ions are allowed to drift or diffuse into the ETL. Fig. \ref{FigModel}b shows band diagrams calculated from a drift-diffusion model corresponding to the various stages illustrated in Fig. \ref{FigModel}a. Two experimental observations favor a mobile, but very slow, new anion species arising from the reduction event. First, $V_\text{br}$ continuously decreases with decreasing $J-V$ scan rate until the $J-V$ scans exceed about 30 minutes in duration (Fig. \ref{Fig:Observations}b, Fig. S2), implying dynamics that require 30 minutes to equilibrate even though the mobile ion concentrations are seen to saturate after only about three minutes (Fig. \ref{Fig_Mobile_Ion_Measurements}b). Second, a new, slower component in the measured current transient appears after reverse bias (Fig. \ref{Fig_Mobile_Ion_Measurements}c, Fig. S10). `\emph{Stage 3}' is illustrated at early times (before new slow mobile anions have completely equilibrated; center panel) and at later times (after complete ionic equilibration; right panel). At early times, the new slow mobile anions have had some time to drift and create a weak concentration gradient (nearly uniform distribution) across the perovskite, enabling steep band bending near the ETL. After complete equilibration, these anions may accumulate into a Debye layer at the ETL interface. Importantly, even with two mobile ion species (cations and anions) at equal concentrations exceeding $\sim$ 10$^{18}$ cm$^{-3}$, charge tunneling barriers at both interfaces become very narrow (nanometers wide) and can still enable high tunneling currents. Tentative timescales for the processes, based on experimental observations, are shown below each stage.  \\ 

The elemental candidates for what gets reduced are Pb cations, A-site cations, and iodine species \cite{kerner_electrochemical_2019, yamilova_reduction_2020, jiang_improved_2024, xu_halogen_2023}. Purely balanced iodide/iodine oxidation/reduction that does not exchange iodine with the lattice would not increase the mobile iodine vacancy concentration and therefore cannot easily explain our observations. A-site cation reduction is unlikely to be the main reaction because reverse-bias degradation can be largely recoverable \cite{jiang_improved_2024,tormena_recoverable_2025}. In this work, the substantial mobile ion recovery shown in Fig. \ref{Fig_Mobile_Ion_Measurements}c,d similarly favors a largely reversible process. A-site cations methylammonium (MA) and formamidinium (FA) are weak Br\o{}nsted acids that can lose their hydrogen ion to participate in a reduction reaction that etches ITO and other metal oxides \cite{kerner_electrochemical_2019, boyd_overcoming_2020}. This process is expected to be irreversible, in part because of the formation of protonated surface oxygen states and hydrogen halide formation that can etch or dissolve the underlying TCO (including ITO), and subsequent migration of indium into the perovskite \cite{kerner_electrochemical_2019,zhan_indium_2024}. We performed time-of-flight secondary ion mass spectroscopy (ToF-SIMS) on cells after about 40 hours of fixed reverse bias ($-3.65$ V) and saw no signal of migrating indium (Fig. S11). We also expect that a high concentration of A-site vacancies and iodine vacancies would promote the perovskite to “structurally collapse” into a PbI$_2$ phase, although XRD measurements after 45 hours of fixed reverse bias ($-3.75$ V) showed no significant sign of PbI$_2$ formation (Fig. S12). Importantly, ToF-SIMS and XRD was performed on cells that showed a very gradual decay in performance (like in Fig. \ref{Fig_Mobile_Ion_Measurements}d) and no signs of thermal damage, so that Joule heating likely cannot contribute to significant structural or chemical changes. The remaining candidate for the reduced species is a Pb cation. Under reverse bias, iodine vacancies --- physically, under-coordinated Pb sites in the perovskite --- are expected to accumulate at the ITO/HTL interface at concentrations locally exceeding 10$^{20}$ cm$^{-3}$ (Fig. S13) \cite{bertoluzzi_mobile_2020, schmidt_characterization_2025}. These sites would be immediately available once the cell is put under reverse bias since they arise from native iodine vacancies. Previous reports indicated that unstable Pb$^{2+}$ termination planes, or under-coordinated Pb, are the most likely sites for charge trapping and corrosion after electron injection \cite{shallcross_determining_2017, birkhold_interplay_2018}. Under-coordinated Pb at the ITO/HTL/perovskite interface due to iodine vacancy accumulation is thus the most likely reduced species if electron transfer can occur between the ITO and perovskite. In this picture, the reduction of under-coordinated Pb enables the paired reduction-oxidation process in the earliest stages, creating more mobile ions. Reported migration barriers of vacancy-assisted Pb transport ($>0.8$ eV) are larger than those for vacancy-assisted iodine transport ($\lesssim0.6$ eV) \cite{eames_ionic_2015,azpiroz_defect_2015} which is a possible explanation for the slower (seconds) mobile ion decay component in Fig. \ref{Fig_Mobile_Ion_Measurements}c (and Fig. S10) and the even slower evolution (over minutes to hours) seen in slow reverse $J-V$ scans in Fig. \ref{Fig:Observations}b. Quantitative comparison of the diffusion time for Pb and iodine is difficult due to the dependence of migration barriers on factors such as grain structure, perovskite composition, and charge states \cite{shao_grain_2016,mcgovern_grain_2021,pering_effect_2022,tyagi_tracing_2025}. Later saturation of the mobile ion concentration can be rationalized once the concentration of neutral (or positively charged) iodine species becomes high enough to enable an iodide/iodine redox shuttle that can sustain electrochemical current in a vacancy-conserving manner \cite{xu_halogen_2023}.  \\

We note that direct experimental observation of the reduced species is expected to be extremely technically challenging. The highest estimated mobile ion concentrations are $\sim1\times10^{18}$ cm$^{-3}$, and so the concentration of reduced species is also expected to be $\sim1\times10^{18}$ cm$^{-3}$, still only about 0.025\% of the total elemental concentration of the material ($\sim4\times10^{21}$ cm$^{-3}$). Such concentrations within the background of the perovskite matrix provide insufficient contrast for techniques like X-ray photoemission spectroscopy or ToF-SIMS to resolve changes in oxidation states or chemical distributions. Additionally, the reduction events occur at the buried interface, introducing another serious technical challenge. \\

The model illustrated in Fig. \ref{FigModel}a highlights the necessity of a reduction reaction (electron transfer to the perovskite) to balance iodide oxidation. The ITO (or other TCO) has a very high concentration of available electrons that, under reverse bias, would like to transfer to the perovskite. However, a good HTL can block electron transfer between the ITO and the perovskite, preventing reduction and thus preventing the corresponding oxidation. If the HTL is absent, too thin, has pinholes, or other electronic leakage pathways, electron transfer will enable rapid reduction and thus indirectly enable iodide oxidation. Therefore, a sufficiently thick and high-quality HTL should limit the $N_\text{ion}$ increase in reverse bias. To test this hypothesis, we measured reverse-bias-induced mobile ion creation in cells intentionally fabricated \emph{without} an HTL. Inferred $N_\text{ion}$ values in HTL-free cells reached but did not exceed $1\times10^{19}$ cm$^{-3}$ (displaced charge of $\sim1\times10^{13}$ cm$^{-2}$) for both $-4.5$ V and $-6$ V (Fig. \ref{FigModel}c), about $10\times$ ($4\times$) larger than in cells with a thin PTAA HTL. (Fig. S8 shows the complete time evolution of current-transient measurements in HTL-free cells.) This concentration may correspond to the highest values measurable with electrical techniques \cite{schmidt_how_2025, schmidt_quantification_2026}. This implies that our estimates from cells with thin PTAA were accurately determined, and that the \emph{actual} $N_\text{ion}$ value in HTL-free cells may be even higher than inferred. Current transients in those HTL-free cells started to develop a time dependence that resembled faradaic current in a one-dimensional electrode-electrolyte system, possibly indicating electrochemical current arising from a high concentration of new mobile ions relaxing and the large ITO-perovskite contact area (Fig. S10). \\

Although $N_\text{ion}$ appears to decrease rapidly after removing the reverse-bias stress (Fig. \ref{Fig_Mobile_Ion_Measurements}d), a practical device or a cell in an outdoor-employed module may need to withstand moderate reverse-bias conditions for many hours or days. If there are slow irreversible electrochemical reactions under reverse bias, cell performance will gradually degrade. We hypothesized that the same electrochemical pathways that enable higher mobile ion concentrations in cells without an HTL will lead to more severe degradation when cells are held for many hours at a moderate reverse bias. High vacancy concentrations over long durations may lead to permanent structural or compositional changes, permanent chemical changes or electrochemical reactions with the underlying TCO, metal oxides, or transport layers \cite{yang_origin_2015, kerner_electrochemical_2019, zhan_indium_2024, thampy_altered_2020, boyd_overcoming_2020, matveeva_electrochemistry_2005, minenkov_monitoring_2024}, or due to complete iodine loss from the system \cite{kim_large_2018, bogachuk_perovskite_2022}. Figure \ref{FigModel}d (top panel) shows the PCE loss in cells held at $-3.85$ V for up to 45 hours. Cells with a typical 5-nm-thick PTAA HTL lost about 25\% relative (4.5\% absolute) PCE over 50 hours while cells without an HTL lost nearly 90\% relative (7\% absolute) PCE within 21 hours. (The starting PCEs between these two cells were very different; about 19\% for cells with PTAA and about 8\% for cells without an HTL.) The reverse current densities flowing through these cells during the fixed voltage holds (lower panel) were initially about 2 to 8 $\mu$A/cm$^2$ for the 0-nm PTAA cell and about 1 $\mu$A/cm$^2$ for the 5-nm-thick PTAA cell. Increased reverse current in the 0-nm PTAA cell may accelerate the degradation, but our model highlights that the reverse current density and mobile ion concentration are coupled; they likely conspire to accelerate the degradation when redox activity is not suppressed. \\ 

\begin{figure}[h!]
      \includegraphics[width=3.5in]{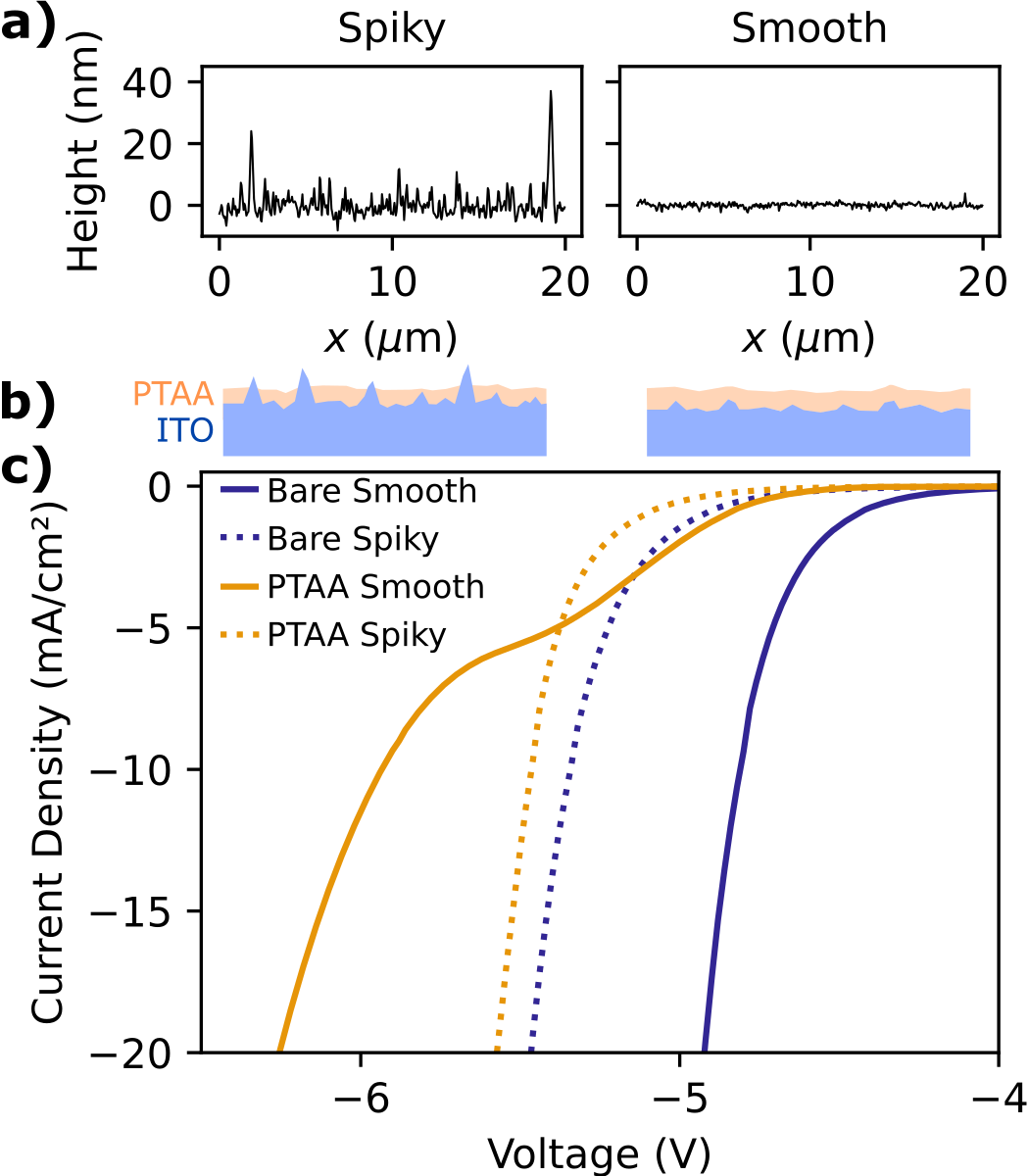}
	\caption{
        (\textbf{a}) Surface texture of ``spiky" ITO (left) and ``smooth" ITO (right) taken from AFM images. 
        (\textbf{b}) Schematic of 5 to 10 nm of PTAA on ``spiky" ITO (left) and ``smooth" ITO (right).
        (\textbf{c}) Measured reverse $J-V$ curves from complete perovskite p-i-n diodes intentionally fabricated without an HTL (blue) and with a 5-nm-thick PTAA HTL (orange). Dashed curves: cells fabricated on ``spiky" ITO. Solid curves: cells fabricated on ``smooth" ITO.
        }
\label{Fig_ITO}
\end{figure}

The impact of ITO-perovskite contact on reverse-bias behavior has implications for manufacturing. Thick PTAA may be able to completely coat the ITO to increase $V_\text{br}$ \cite{jiang_improved_2024}, but it has unacceptably high series resistance and leads to poor PCEs. Thin HTLs such as PACz interface modifiers or thin PTAA used in high-performance devices have known coverage issues \cite{sun_niox-seeded_2021,phung_enhanced_2022,jiang_towards_2023,fei_strong-bonding_2024,contreras_deposition-dependent_2025,qi_disaggregation_2026}, partially due to ITO surface roughness, protrusions, and surface chemistry heterogeneity \cite{jonda_surface_2000, tak_criteria_2002, betz_synthesis_2008}, and are known to be suboptimal for stopping electron transfer \cite{kong_phosphonic-acid-reinforced_2026} which may largely explain smaller $V_\text{br}$s in cells fabricated with those materials. To illustrate the problem, we studied devices fabricated on ITO from two different manufacturers with different preparation procedures to create very different ITO surface textures. Fig. \ref{Fig_ITO}a shows surface height profiles from atomic-force microscopy (AFM). (Full AFM images and further surface analyses are shown in Fig. S17). One substrate has a high concentration ($>3$ spikes per $\mu$m$^2$) of distinct protrusions (``spikes") up to 40 nm tall. The other texture has very few ($<0.1$ spikes per $\mu$m$^2$) spikes, and almost none taller than 20 nm. We will refer to these as ``spiky" and ``smooth" ITO, respectively. Cyclic voltammetry measurements indicate that thin PTAA prepared on ``spiky" ITO leaves a large area of exposed ITO, while the same PTAA deposition procedure coats the ``smooth" ITO very well --- although not perfectly --- and substantially limits electron transfer between ITO and electrolyte (Fig. S18) \cite{minenkov_monitoring_2024}. Fig. \ref{Fig_ITO}b illustrates the coverage of PTAA on these two ITO textures. In a complete perovskite p-i-n diode, exposed ITO will directly contact perovskite, enabling electron transfer and electrochemistry between those layers \cite{shallcross_determining_2017, kerner_electrochemical_2019, minenkov_monitoring_2024}. Fig. \ref{Fig_ITO}c shows measured reverse $J-V$ curves from complete cells fabricated on these ITO/PTAA contacts (orange curves). Cells fabricated on ``smooth" ITO (solid curves) showed the highest $V_\text{br}$ values. We also fabricated and measured reference structures \emph{without} the HTL (blue curves). On ``spiky" ITO, the PTAA layer only weakly affected the reverse $J-V$ curve, while on ``smooth" ITO, the PTAA layer clearly increases $V_\text{br}$. This result indicates that the HTL critically impacts reverse-bias behavior primarily due to its ability to cover the ITO surface and block reduction reactions. $V_\text{br}$ is increased only when the HTL sufficiently covers the ITO.\\

To explore the ability of a metal oxide layer to slow electron transfer and the resulting mobile ion concentration increase, we fabricated cells with 20-nm-thick nanoparticle NiO$_x$ HTLs capped with 5-nm-thick PTAA or phosphonic-acid carbazole (PACz) `self-assembled monolayer' (SAM) HTLs. Fig. S14 shows that these HTLs did not stop the increase in mobile ion concentration under reverse bias, likely due to poor coverage or ion-blocking capabilities in the case of SAMs\cite{phung_enhanced_2022} and the high concentration of electroactive surface states present on NiO$_x$ \cite{boyd_overcoming_2020,xie_electrochemical_2026} (like many other metal oxides \cite{yang_origin_2015, thampy_altered_2020}) and the very high surface areas of nanoparticle films. We also explored fluorine-doped tin oxide (FTO) TCOs with PACz SAMs HTLs. All combinations show the same qualitative trend of increasing $N_\text{ion}$ with time under reverse bias with typical maximum values between $10^{18}$ cm$^{-3}$ and $10^{19}$ cm$^{-3}$. Cells with PTAA on ITO show the smallest $N_\text{ion}$ increase. We believe that combining \emph{conformal} deposition (ALD or sputtering) of dense metal oxides for good electron blocking and suppression of defect-driven reverse-bias behavior \cite{amorales_nickel-oxide_2026} with thin capping layers such as PACz/polymer combinations \cite{zhao_ambient_2025} or, ideally, tunnel-oxide layers to passivate the oxide surface and block ions may be a promising route for stabilizing transport materials. \\

\section*{DISCUSSION}
We used reverse $J-V$ measurements, charge-tunneling calculations and mobile ion current-transient measurements to verify that reverse $J-V$ behavior and evolution in perovskite p-i-n diodes is dominated by mobile ions created under reverse bias. New mobile cations (iodine vacancies) are created when iodide is oxidized in the cell under reverse bias. Mobile ion concentrations of about $1\times10^{18}$ cm$^{-3}$ are created within just three minutes under $-6$ V in p-i-n cells fabricated with a 5-nm-thick PTAA HTL and thick SnO$_x$. Cells intentionally fabricated without an HTL showed about $10\times$ larger mobile ion concentrations ($1\times10^{19}$ cm$^{-3}$) under these same stress conditions. Importantly, since the role of an HTL is to block electron transfer, this verifies that oxidation reactions are enabled only when reduction reactions are enabled between the bottom electrode (ITO or any other TCO) and perovskite. These mobile ion concentration dynamics explain low $V_\text{br}$ values, why $V_\text{br}$ decreases over time under reverse bias, and even the precise shape of measured reverse $J-V$ curves. \\

This study illustrates that perovskite diodes are both semiconductor devices and electrochemical cells. Understanding the semiconductor device behavior requires an accurate band diagram that can only be calculated if the impact of electrochemical reactions on mobile ion concentrations is considered. Although many previous studies have focused on ETL engineering \cite{li_barrier_2024, ren_mobile_2024, lanzetta_tinlead_2025}, HTL engineering will be key to stabilizing perovskite cells under reverse bias \cite{jiang_improved_2024, zhao_ambient_2025}. Designing conformal HTLs to rigorously block both electron transfer and ion migration will shut off reduction reactions between the ITO and perovskite. The mobile ion concentration will not increase significantly under reverse bias, leading to increased $V_\text{br}$ values and minimized degradation. The damage from electrochemical reactions may not be as severe as some might expect because it appears that a steady state equilibrium of about $1\times10^{18}$ cm$^{-3}$ is quickly reached in the film. Still, at this concentration, degradation will occur over many hours or days even if irreversible reaction rates are slow. If iodine permanently reacts with a transport layer or the electrode, or if it leaves the device, then permanent degradation will certainly occur. A good ETL with excellent ion-blocking is critical to ensure that iodine stays in the perovskite. One option is to grow SnO$_x$ directly on perovskite \cite{gao_long-term_2024}. If iodine does not leave the cell and if other permanent chemical changes are inhibited, iodine may eventually reincorporate into the lattice, and full cell recovery may be possible. While we have focused this study on reverse bias, changes in mobile ion concentrations are expected to also be important under other operating conditions \cite{thiesbrummel_ion-induced_2024}. This study therefore has implications for stability in forward bias, especially for light-emitting diodes that are operated at relatively large voltage and current densities.




\section*{METHODS}

\section{Fabrication Methods}

\subsection{Materials}
\subsubsection{Perovskite composition}
The primary perovskite composition used in this work was Cs$_{0.25}$FA$_{0.75}$Pb(I$_{0.85}$Br$_{0.15}$)$_3$ with a band gap of approximately 1.65 eV. In Figs. S2, S3, and S14 different perovskite compositions were used to test the generality of our observations.

\subsubsection{Perovskite precursor materials}
Lead iodide (``PbI$_2$", $\ge$ 99.99\% (trace metal basis)) was purchased from TCI. Cesium iodide (``CsI") was purchased from Sigma Aldrich. Lead (II) bromide (``PbBr$_2$", Puratronic $\ge$ 99.998\% (trace metal basis)) was purchased from Thermo-Fischer Scientific. Formamidinium iodide (``FAI") ($\ge$ 99.99\%) was purchased from GreatCell Solar Materials. \\

\subsubsection{Hole transport materials}
Poly[bis(4-phenyl)(2,4,6-trimethylphenyl)amine] (``PTAA", 11,000 kDA) was purchased from Ossila. \\

\subsubsection{Electron transport materials}
C$_{60}$ fullerene ($\ge$ 99.9\% sublimated) was purchased from Sigma Aldrich. Bathocuproine (sublimated) (``BCP", $\ge$ 99.9\%) was purchased from TCI Chemicals. Tetrakis-dimethylamino tin (IV) (``TDMA-Sn" for ALD SnO$_x$, 99.99\% Sn basis) was purchased from Strem Chemicals. \\

\subsubsection{Top contact}
Silver (Ag) pellets were purchased from Kurt J. Lesker. \\

\subsection{Fabrication}

\subsubsection{Solvent preparation}
Chlorobenzene  (CB) ($>$99.9\% anhydrous), Ethyl Alcohol (Et-OH)  ($>$99.9\% anhydrous), N,N-Dimethylformamide (DMF) ($>$99.8\% anhydrous), and 1-Methyl-2-pyrrolidone (NMP) ($>$99.5\% anhydrous) were purchased from Sigma-Aldrich. All solvents were filtered through a 0.2 $\mu$m PTFE filter prior to use.

\subsubsection{Substrate Preparation} 
Tin-doped indium Oxide (ITO)-coated glass substrates (25 mm $\times$ 25 mm) were purchased from Delta Technologies and Thin Film Devices.``Smooth" ITO used substrates from Thin Film Devices that were brushed vigorously with a nylon-bristle toothbrush for about 1 minute per substrate in deionized water.``Spiky" substrates were used as received from Delta Technologies. All substrates (smooth and spiky) were then rinsed under tap water, then DI water to clean off particles. Each substrate was then cleaned by ultrasonication in DI water, acetone, and isopropanol for 15 minutes each. Substrates were removed and dried with a stream of nitrogen, then subjected to a 20 minute UV-Ozone treatment immediately prior to fabrication. \\

\subsubsection{HTL fabrication} 
For devices using PTAA, the PTAA was mixed at 1.5 mg/mL in CB for at least 24 hours prior to fabrication using a stir bar on a stir plate. For 50 nm of PTAA, we used 15 mg/mL. 70 $\mu$L of PTAA solution was deposited on a static substrate and spun at 5,000 rpm for 20 seconds (2500 rpm/s acceleration), then annealed on a hotplate at 100$^\circ$C for 10 minutes.

\subsubsection{Perovskite fabrication}
The perovskite solution was created by dissolving 0.0747 g of CsI, 0.0950 g of PbBr$_2$, 0.1483 g of FAI, and 0.4109 g of PbI$_2$ for each 1 mL of desired solution to create a 1.15 molar solution of Cs$_{0.25}$FA$_{0.75}$Pb(I$_{0.85}$Br$_{0.15}$)$_3$. The solvent was mixed 6:1 DMF:NMP by volume, and was mixed prior to adding to the perovskite precursors. After adding solvents to the perovskite precursors, the solution was mixed using a stir bar on a stir plate for approximately 24 hours before fabrication. Approximately 30 minutes prior to fabrication, the solution was passed through a 0.2 $\mu$m PTFE filter. \\

After annealing the HTL, substrates were allowed to cool (approximately 15 minutes), then the perovskite layer was fabricated. 200 $\mu$L of perovskite solution was deposited on a static substrate and spun in a 4-step process. Step 1: 2,000 rpm for 10 seconds (1,000 rpm/s acceleration). Step 2: 6,000 rpm for 7 seconds (2,000 rpm/s acceleration). Step 3: 6,000 rpm for 25 seconds with $N_2$ gas quenching. Step 4: 6,000 rpm for 7 seconds. Films were then annealed at 100 $^\circ$C for 30 minutes. \\

The gas quenching setup is described in prior work \cite{kaczaral_improved_2023}.

\subsubsection{ETL and top contact fabrication} 
30 nm of C$_{60}$ was deposited in an Angstrom Engineering thermal evaporator at a pressure of $\sim 5 \times10^{-7}$ Torr at rates between 0.1 and 0.3 \AA/s. Approximately 30 nm of SnO$_x$ was deposited by atomic-layer deposition (ALD) in a Beneq TFS-200 ALD system by reacting TDMA-Sn and de-ionized water at a reactor temperature of about 90$^\circ$C. The TDMA-Sn precursor vessel was heated to about 55$^\circ$C and the DI water was unheated. TDMA-Sn was dosed by backfilling the bubbler with nitrogen for 350 ms and then dosing for 250 ms. Growth cycles proceeded by: [TDMA-Sn dose (350 ms charge/250 ms dose) - purge (5 s) - DI water dose (150 ms) – purge (3 s)]. 250 cycles were used to deposit approximately 30 nm of SnO$_x$. For the top contact in ``full-size" (12 mm$^2$) cells, 120 nm of silver was thermally evaporated at a pressure of $\sim 5\times10^{-7}$ Torr at rates between 0.1 and 1.5 \AA/s. Masks were used to create six square devices on each substrate. Samples were rotating during the metal deposition. For the 0.0314 mm$^2$ (200-micrometer diameter) cells (``micro cells"), 300 nm of Ag as evaporated at a pressure of $\sim 5\times10^{-7}$ Torr at rates between 0.1 and 1.5 \AA/s using a mask that contained a 5 $\times$ 25 array of 200 $\mu$m diameter circles. Samples were not rotated during the Ag deposition for these micro devices. $J-V$ and current transient measurements shown in Fig. \ref{Fig:Observations}, and Fig. \ref{Fig_Mobile_Ion_Measurements} were recorded from micro cells. PCE degradation measurements and $J-V$ measurements shown in Figs. \ref{FigModel} and \ref{Fig_ITO} were recorded from 12 mm$^2$ cells. \\

\section{Experimental Methods}
\subsection{Current-voltage and current transient measurements}
Electrical measurements ($J-V$ curve measurements and current transient measurements) were performed using a Keithley 2400 sourcemeter for both applying external potentials to our cells and for measuring the current signal from the cells. All measurements were performed in an N$_2$-filled glovebox. For measurements on micro (200-$\mu$m diameter) cells, the top electrodes were contacted using a thin ($\sim 100$ $\mu$m diameter) copper wire probe that was connected to a ``micromanipulator" probe station with a microscope to enable selective electrical contact to single specific cells. The ground probe was contacted directly to the ITO substrate, typically millimeters from the tested cell to minimize series resistance. Electrical contact between the probe and cell was confirmed by applying a small positive voltage across the probes and slowly lowering the top-contact probe until a stable current was measured. \\

For reverse $J-V$ measurements, the applied potential was swept from 0 V to increasingly negative potentials while the current density was recorded at each step. For fast sweeps ($\geq$1 V/s), the Keithley sourcemeter's internal buffer mode was used. For slower sweeps ($<$ 1 V/s), the current was read out at each applied voltage step. \\ 

Mobile ion concentrations were estimated from measurements based on the experimental technique developed and described in Ref. \cite{bertoluzzi_situ_2018} without optical pumping. The mobile ion concentration measurement procedure involves two steps. First, a potential $V_1$ is rapidly applied to the cell at time $t=t_0$. That potential is maintained for a duration $t_1$, over which ions are ``pushed" to one side of the cell. The applied potential is then abruptly changed to a potential $V_2$. In our experiments, $V_2$ is always 0 V. Mobile ions in the perovskite equilibrate to this new potential. Electronic charges in the external circuit respond to this internal ion motion, providing a decaying current transient signal that can be measured with an external ammeter. In the time range $t>t_0+t_1$ (when the applied potential is 0 V) the current will be almost purely the ionic displacement current. Integrating the current density over time directly gives the total displaced \emph{external electronic} charge during ion equilibration. The total displaced electronic charge is directly related to the mobile ion concentration and the initial applied potential $V_1$. The relationship is plotted in Fig. S9 over a range of mobile ion concentrations and for a selection of applied potentials ($+0.8$ V and negative potentials). Complete mathematical details are provided in Supplementary Information Section 6. In all reported values in this work, the first $\approx 20$ ms this final transient was intentionally omitted to avoid residual electronic contributions from charge extraction or injection into the cell. The electronic portion (first 20 ms) of the current transient amounts to displaced charge values of about $3\times10^{11}$ cm$^{-2}$ for $V_1=-3.0$ V and about $1\times10^{12}$ cm$^{-2}$ for $V_1=-6.0$ V. Curves were integrated out to 8 s. Current density values below a specified noise level of 0.1 $\mu$A/cm$^2$ were omitted from the integration. \\

Measurements were performed using the Keithley sourcemeter's internal buffer mode in order to minimize the sampling period. Current ``integration times" were set by the number of power line cycles (`NPLC') to 1 to achieve the lowest noise possible in each current reading. This corresponds to sampling times of approximately 16.67 ms and sampling periods (time between successive current readings) as low as 20 ms. Current-reading noise levels were regularly below $4\times10^{-2}$ $\mu$A/cm$^2$, significantly smaller than the ionic current transient signals (typically 0.1-10 $\mu$A/cm$^2$). The $V_1$ duration $t_1$ was varied between about 60 ms to 300 s. $V_1$ values of about $-2$ V to $-6$ V were tested. After switching to 0 V, the current was recorded for 8 seconds. \\

\subsection{Cyclic voltammetry measurements}
Cyclic voltammetry was performed using a solution of methylammonium iodide in dimethylformamide at a solution concentration of 0.3 M. 1-inch square patterned ITO substrates (with and without HTLs) were submersed about 1 cm deep into this solution. The ITO electrode was taken through one CV cycle from 0 V to $-1.55$ V and back to 0 V, relative to an Ag/AgCl reference electrode, at a scan rate of about 50 mV/s. The optical transmittance spectrum through the samples was recorded continuously during the CV cycle. Optical transmittance spectra at each time step was normalized to the initial optical transmittance spectrum before the CV cycle. That is, the transmittance is equal to 1 at all wavelengths before the CV cycle. The transmittance spectra and optical images shown in the manuscript are the final transmittance spectra after the cycle was complete. Electrochemical reactions between the ITO and electrolyte are further evidenced by an advanced yellowing of the solution, likely due to a high concentration of iodine or triiodide in the final solution.

\subsection{Fixed reverse potential stability measurements}
Reverse-bias stability measurements were performed in a Fluxim LitosLite multi-channel stability tester. The measurement routine involved the following three steps comprising a sequence that was indefinitely repeated: 1) maximum-power-point tracking (MPPT) for 6 minutes under 0.85 sun global intensity with a nominal AM1.5G simulated solar spectrum; 2) measurement of forward and backward light-JV curve under the same simulated spectrum; 3) dark fixed reverse potential stress at -3.85 V for 35 minutes. Figure 4d in the manuscript plots the cell PCE as a function of the cumulative reverse-bias stress duration. Current densities are continuously recorded during the fixed reverse potential stress steps with a point rate of 12 1/min. Samples were not actively heated or cooled during this campaign. 

\subsection{Time-of-flight secondary ion mass spectroscopy (TOF-SIMS)}
An ION-TOF TOF-SIMS V Time of Flight SIMS (TOF-SIMS) spectrometer was utilized for depth profiling utilizing methods covered in detail in previous reports \cite{harvey_investigating_2020}. Analysis was completed utilizing a 3-lens 30 keV BiMn primary ion gun. High mass resolution depth profiles were completed with a 30 KeV Bi$^{3+}$ primary ion beam, (0.8 pA pulsed beam current), a $50\times50$ $\mu$m area was analyzed with a 128:128 primary beam raster. Sputter depth profiling was accomplished with 1 kV Cesium ion beam (6 nA sputter current) with a raster of $150\times150$ $\mu$m. TOF-SIMS was performed on cells after 40 hours of fixed reverse bias at $-3.85$ V. Only cells that did not show signs of abrupt breakdown or electrical shorting during the fixed reverse bias were measured by TOF-SIMS. TOF-SIMS measurements were performed about 3-4 days after the fixed reverse bias tests were complete. All cells first had the top metal contact mechanically delaminated by tape before TOF-SIMS measurements. An ALD SnO$_x$ layer that was originally between the C$_{60}$ and top silver contacts evidently was removed during the delamination, as there were no Sn or O signals above the C (C$_{60}$) signal.


\newpage


\section*{RESOURCE AVAILABILITY}


\subsection*{Lead contact}


Requests for further information and resources should be directed to and will be fulfilled by the lead contact, Michael D. McGehee (michael.mcgehee@colorado.edu).

\subsection*{Materials availability}


This study did not generate new materials.

\subsection*{Data and code availability}


\begin{itemize}
    \item Data, original codes, and simulation support files required to reproduce the analyses in this article are available from Mendelay Data and DOI: 
    \item Any additional information required to reanalyze the data reported in this paper is available from the lead contact upon request.    
\end{itemize}

\section*{ACKNOWLEDGMENTS}


This material is based upon work supported by the U.S. Department of Energy’s Office of Energy Efficiency and Renewable Energy (EERE) under Solar Energy Technologies Office (SETO) Agreement Number DE-EE0010502. This work was authored in part by the National Laboratory of the Rockies for the U.S. Department of Energy (DOE) under Contract No. DE-AC36-08GO28308. D.G. and F.J. acknowledge U.S. Department of Energy, Office of Basic Energy Sciences, Division of Materials Sciences and Engineering under Award DE-SC0013957 to support their writing and analysis. The views expressed in the article do not necessarily represent the views of the DOE or the U.S. Government. We thank Kaushik Jayaram and Alex Hedrick from the University of Colorado College of Engineering and Applied Science for providing micromachined evaporation masks that enabled us to perform measurements on miniature solar cell devices.

\section*{AUTHOR CONTRIBUTIONS}


Conceptualization: K.F., R.D., D.M., M.M.
Data curation: K.F., R.D., M.P.
Formal Analysis: R.D., K.F.
Funding Acquisition: J.L., M.M.
Investigation: K.F., R.D., M.P., K.S.
Methodology: K.F., R.D., X.Z., Y.Y., B.C., K.S.
Project Administration: R.D., M.M.
Resources: E.S., D.G., J.L., B.C., M.M.
Supervision: R.D., J.L., N.A., M.M.
Visualization: K.F., R.D.
Writing - Original Draft Preparation: K.F., R.D., M.M.
Writing - Review \& Editing: All authors.
\section*{DECLARATION OF INTERESTS}


McGehee is an advisor to Swift Solar.



\section*{SUPPLEMENTAL INFORMATION INDEX}




\begin{description}
  \item Document S1. Supplemental discussion, methods, and Figures S1-S18
\end{description}

\newpage

\bibliography{references}

\bigskip


\renewcommand{\thefigure}{S\arabic{figure}}
\setcounter{figure}{0}
\renewcommand*{\thesection}{\arabic{section}}
\setcounter{section}{0}

\newpage

\newcommand{\ban}[1]{\begin{align*}#1\end{align*} }
\newcommand{\ba}[1]{ \begin{align}#1\end{align} }
\newcommand{\half}{\frac12}
\newcommand{\thalf}{\frac32}
\newcommand{\overbar}[1]{\mkern 1.5mu\overline{\mkern-1.5mu#1\mkern-1.5mu}\mkern 1.5mu}

\section*{Supplemental Information: Electrochemical reactions under reverse bias create additional mobile ions that enable hole tunneling in metal halide perovskite diodes}

\section{Manuscript-referenced Supplementary Figures}

\begin{figure}[h!]
    \centering
    \includegraphics[width=5in]{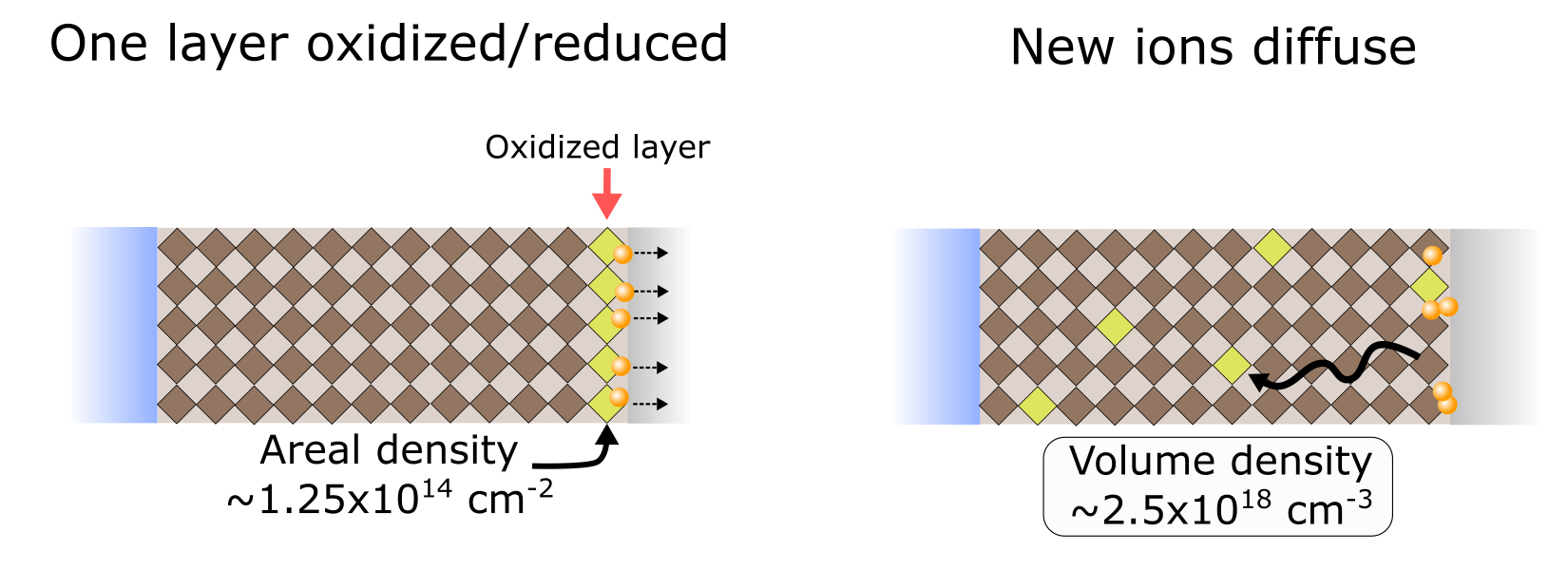}
	\caption{
    Order-of-magnitude estimate of mobile ion concentrations generated under oxidizing conditions. This calculation is detailed in a section of this Supplemental Information document.
    }
\label{SI_Layer_of_Oxidation}
\end{figure}

\begin{figure}[h!]
    \centering
    \includegraphics[width=3in]{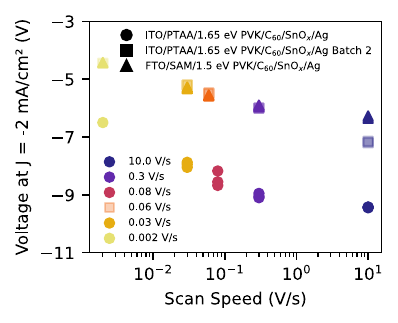}
	\caption{
    Breakdown voltage of perovskite diodes derived from reverse $J-V$ curves as a function of reverse potential scan speed. Results from three different samples are shown, each from different batches fabricated on different days and in two different labs. All show the same trend of decreasing breakdown voltages for slower scans. 1.65-eV-bandgap devices on tin-doped indium oxide (ITO) were fabricated at University of Colorado Boulder. 1.55-eV-bandgap devices on fluorine-doped tin oxide (FTO) were fabricated at Northwestern University.
    }
\label{SI_All_Speed_Dependent_ScanToJ}
\end{figure}

\begin{figure}[h!]
    \centering
    \includegraphics[width=3in]{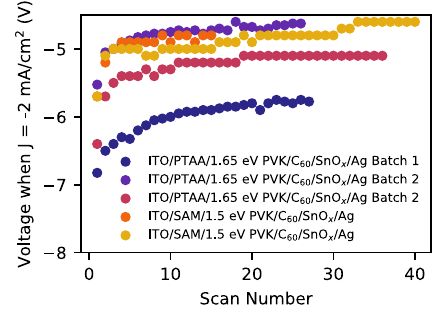}
	\caption{
    We performed repeated scans (on the same device) at a constant scan rate and plotted the progression of the breakdown voltage versus scan number. Each sample across 3 different batches shows the same trend of decreasing breakdown voltage with scan number. 1.65-eV-bandgap cells were fabricated at the University of Colorado. 1.55-eV-bandgap cells were fabricated at Northwestern University. This shows that across different architectures, breakdown voltage still decreases (moves towards 0 V) as the device spends longer under reverse bias.
        }
\label{SI_AllRepeat_ScanToJ}
\end{figure}

\begin{figure}[h!]
    \centering
    \includegraphics[width=7in]{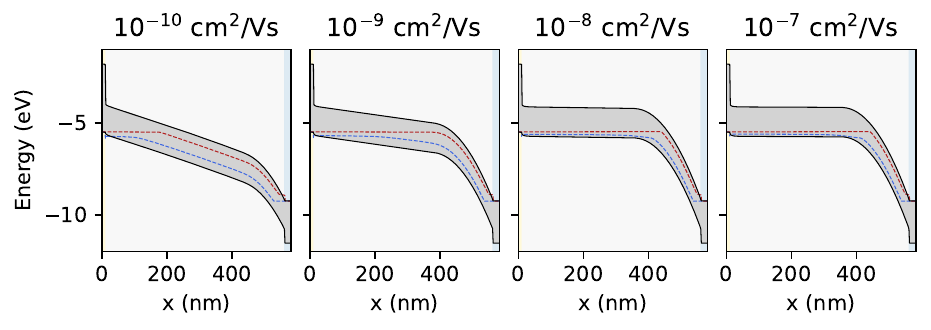}
	\caption{Band diagrams calculated using a drift-diffusion model with mobile cations at a constant concentration of $1\times10^{17}$ cm$^{-3}$. Energy levels are shown at $-5$ V during a voltage sweep from 0 V to $-5$ V at 10 V/s. Each panel is calculated for a different cation mobility (specified above each panel). At a 10 V/s sweep rate, a mobility $>10^{-9}$ cm$^2$/Vs is expected to be sufficiently high to enable complete ionic equilibrium during the sweep. Alternatively, at a cation mobility $10^{-9}$ cm$^2$/Vs, sweep rates $<$10 V/s are expected to be sufficiently slow to enable complete ionic equilibrium during the sweep.
        }
\label{Fig_SI_Drift_Diffusion_Calcs}
\end{figure}

\begin{figure}[h!]
    \centering
    \includegraphics[width=3in]{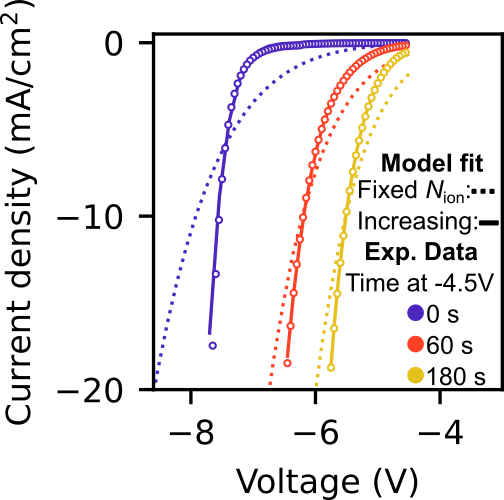}
	\caption{Open circle markers: Experimental reverse $J-V$ scans measured with a scan rate of 4 V/s after holding $-4.5$ V across the cell for 0 s (blue), 60 s (red), and 180 s (gold). Dotted and solid curves: Calculated $J-V$ curves using a hole-tunneling model with two different assumptions for the mobile ion concentration $N_\text{ion}$. Dotted curves: fixed $N_\text{ion}$. Solid curves: linearly increasing $N_\text{ion}$. The fixed $N_\text{ion}$ values needed to best fit each experimental curve are 2.3$\times10^{18}$ cm$^{-3}$ (0 s hold time), 2.9$\times10^{18}$ cm$^{-3}$ (60 s hold time), and 3.3$\times10^{18}$ cm$^{-3}$ (180 s hold time). The varying $N_\text{ion}$ value are presented in the manuscript.
    }
\label{}
\end{figure}

\begin{figure}[h!]
    \centering
    \includegraphics[width=5.5in]{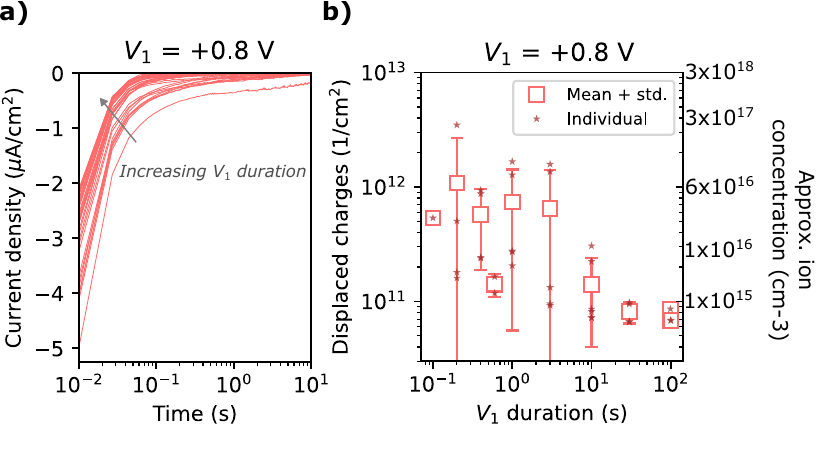}
	\caption{
    (\textbf{a}) Raw current vs. time data immediately after switching to 0 V after holding $V_1 = 0.8$ V for varying durations between 100 ms and 100 s.
    (\textbf{b}) Measured displaced electronic charge and inferred mobile ion concentrations in a perovskite p-i-n diode with a 5-nm-thick PTAA HTL after holding $V_1 = 0.8$ V for varying durations between 100 ms and 100 s. 
    }
\label{}
\end{figure}

\begin{figure}[h!]
    \centering
    \includegraphics[width=3.5in]{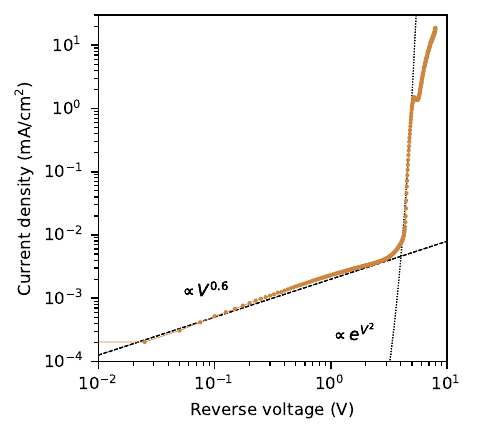}
	\caption{Reverse current density vs. voltage ($J-V$) curve of a perovskite p-i-n diode fabricated with 5-nm-thick PTAA, presented on a log-log scale to highlight $J-V$ scaling behavior. Tunneling current is identified as very steep $J-V$ scaling behavior (locally consistent with $\propto\exp(V^2)$) for reverse potentials $\geq|-4|$ V. 
    }
\label{}
\end{figure}

\begin{figure}[h!]
    \centering
    \includegraphics[width=7in]{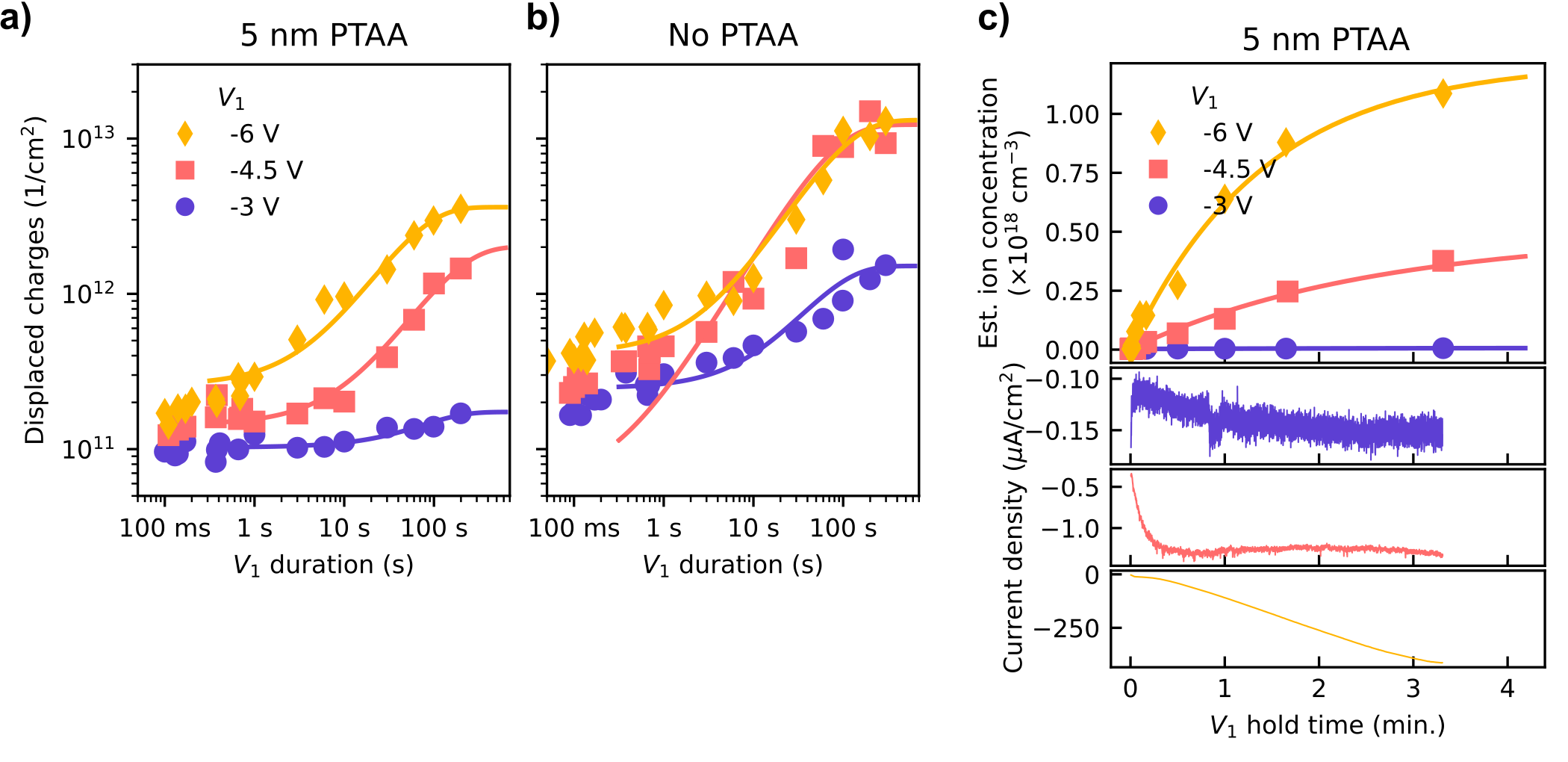}
	\caption{
    (\textbf{a-b}) Measured external integrated charge from ionic current transient measurements in perovskite diodes fabricated with 5-nm-thick PTAA (a) and 0 nm PTAA (b) as a function of reverse-bias $V_1$ hold duration ($x$ axis) for three different $V_1$ values: $V_1=-3.0$ V (blue), $V_1=-4.5$ V (red), and $V_1=-6.0$ V (gold). The data (markers) plotted in (a) is the same as that in the manuscript. Data is presented on a log-log scale to emphasize timescales and absolute changes in ion concentrations. 
    (\textbf{c}) Mobile ion concentration evolution plotted on a linear-linear scale (top panel) with the corresponding typical measured current densities during the fixed $V_1$ hold (bottom three panels) for the three applied potentials: $-3$ V (blue, first row), $-4.5$ V (red, second row), and $-6$ V (gold, third row). Only a single current-density curve is shown for each potential, corresponding to the longest hold of 200 s. Mobile ion concentration values are derived by scaling the displaced charge in (a) according to the displaced ion model described in Section 6. 
    }
\label{}
\end{figure}

\begin{figure}[h!]
    \centering
    \includegraphics[width=0.8\textwidth]{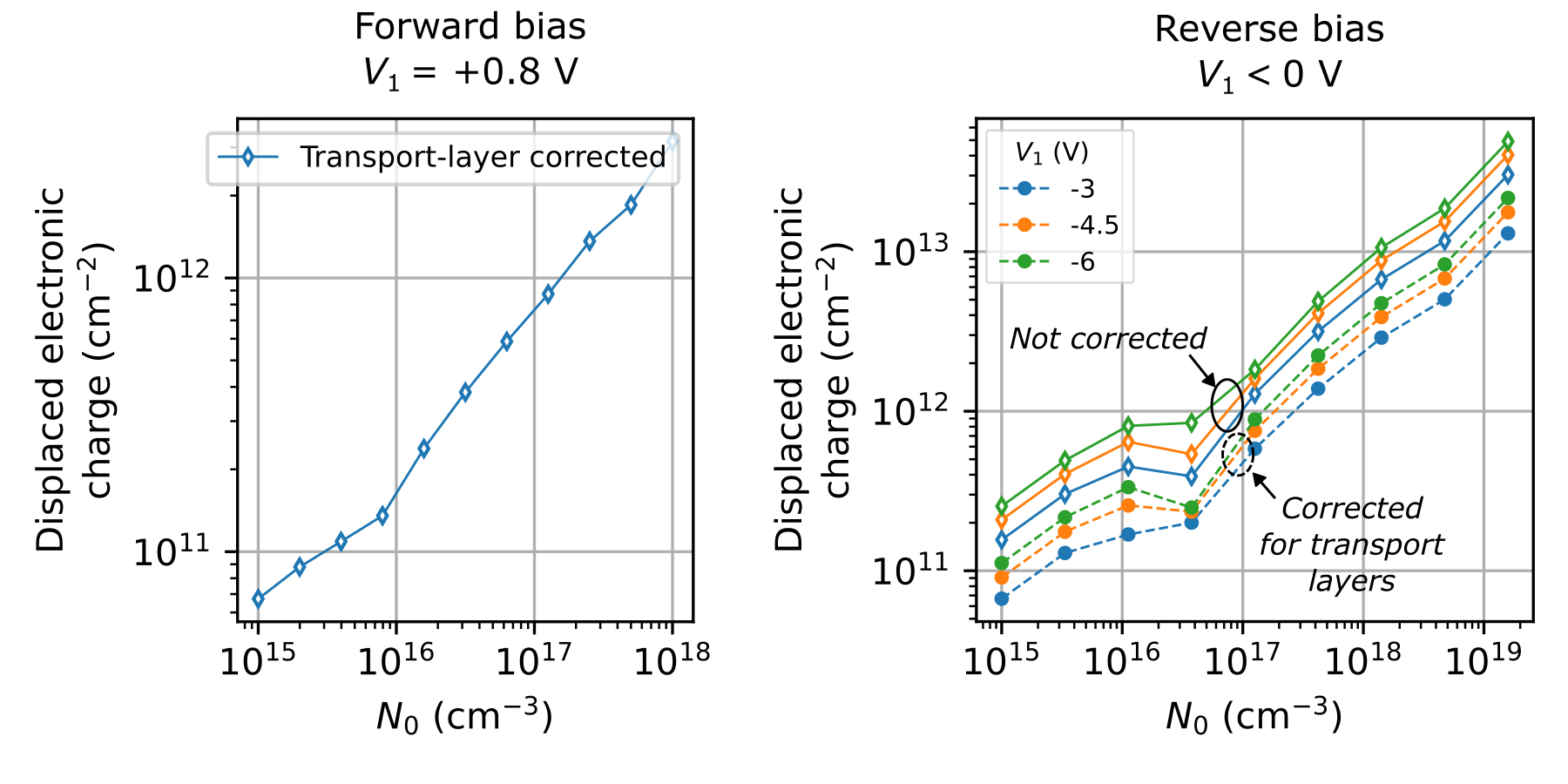}
	\caption{
    Numerically calculated displaced charge vs. $N_0$ for a forward potential $V_1=+0.8$ V (left panel) and reverse potentials $V_1<0$ V (right panel). For reverse potentials, we calculate curves for $V_1$ = -3 V (blue), -4.5 V (orange), and -6.0 V (green). Solid curves assume the applied potential $V_1$ is completely dropped across the perovskite layer. Dashed curves approximately account for potential variations across the transport layers (mostly the ETL since the HTL is very thin in this work) using the model described in this SI Section. 
    }
\label{SIfig:Displaced_Charge_vs_N0}
\end{figure}

\begin{figure}[h!]
    \centering
    \includegraphics[width=7in]{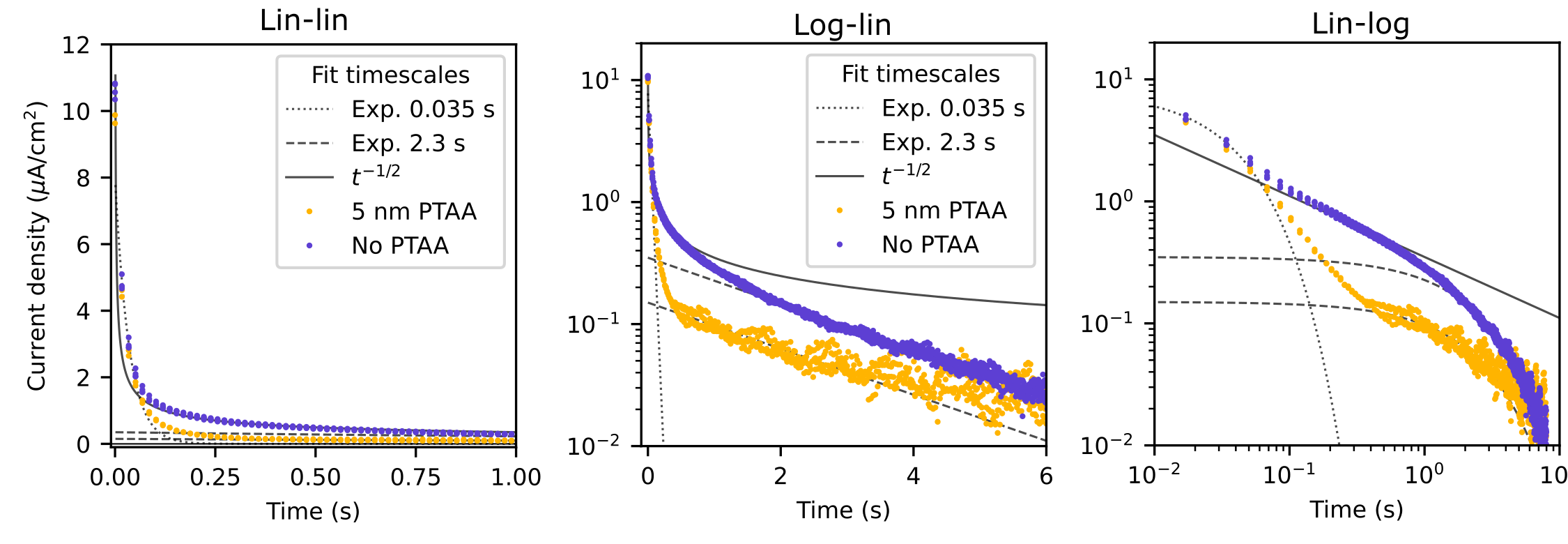}
	\caption{Current transients immediately after switching to 0 V from $-6$ V for 200 s in cells with 5-nm-thick PTAA (gold) and for 300 s in cells with no HTL (no PTAA) (blue). The same data and theoretical curves are plotted with three different x- and y-scale combinations (specified above each plot). Three independent theoretical decay functions are plotted over the data. Black dotted curve: mono-exponential decay with decay time constant of 35 ms. Black dashed curves: mono-exponential decay with decay time constant of 2.3 s (two curves are shown with different coefficients: 0.35 $\mu$A/cm$^2$ for no PTAA and 0.25 $\mu $A/cm$^2$ for 5-nm-thick PTAA). Black solid curve: $J(t)\propto t^{-1/2}$, which would represent diffusion-limited faradaic current in a one-dimensional electrode-electrolyte system. The 5-nm-thick PTAA cell decay appears to be dominated by exponential decay behavior. The no-PTAA cell appears to develop a new characteristic time dependence that is reasonably well represented by $\propto t^{-1/2}$ behavior, at least between about 100 ms and 1 s.}
\label{Fig_Transient_w_Fits}
\end{figure}

\begin{figure}[h!]
    \centering
    \includegraphics[width=6in]{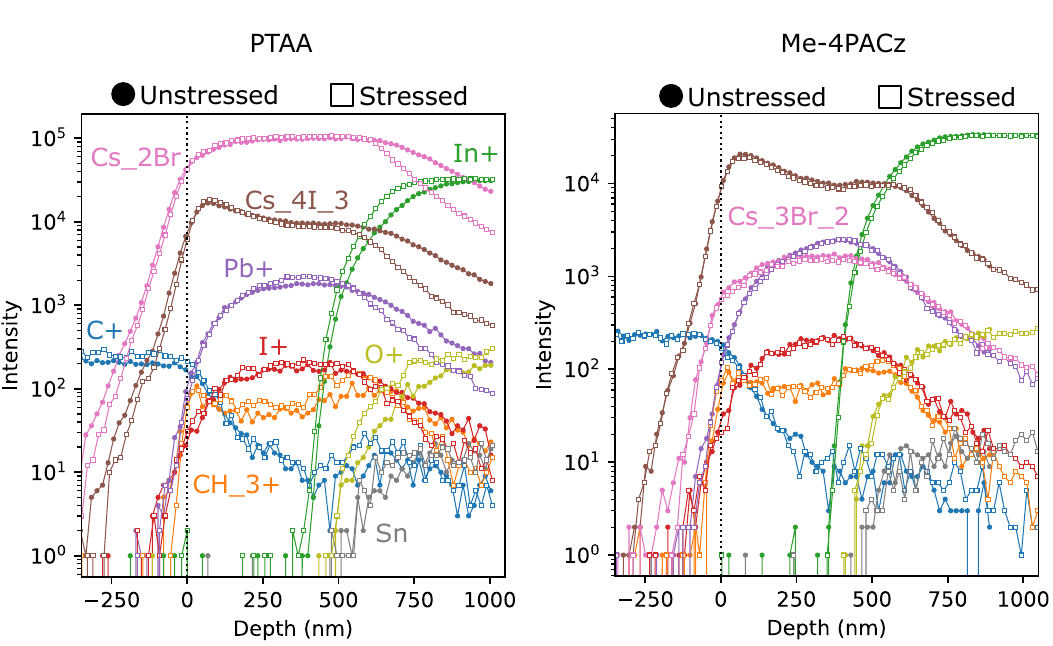}
	\caption{Time-of-flight secondary ion mass spectroscopy (ToF-SIMS) data showing elemental distributions across the depth of samples without reverse-bias stress (solid circle markers) and after (open square markers) fixed reverse bias ($-3.65$ V) for 75 hours. Elements or complexes are distinguished by color and specified by colored text near each trace. Left panel: cells fabricated with PTAA HTL on ITO. Right panel: Cells fabricated with Me-4PACz HTL on ITO. TOF-SIMS data was performed only on cells that did not show abrupt failure and shorting during the long-term fixed reverse bias stress test. The depth ($x$-axis) was artificially scaled so that the C$_{60}$-perovskite interface is roughly at $x$=0 and the perovskite-HTL-ITO interface is roughly at 550 nm (the thickness of our perovskite films). TOF-SIMS data on stressed cells show no migration of the In$^+$ signal into the perovksite compared to unstressed cells. 
    }
\label{Fig_TOF-SIMS}
\end{figure}

\begin{figure}[h!]
    \centering
    \includegraphics[width=5in]{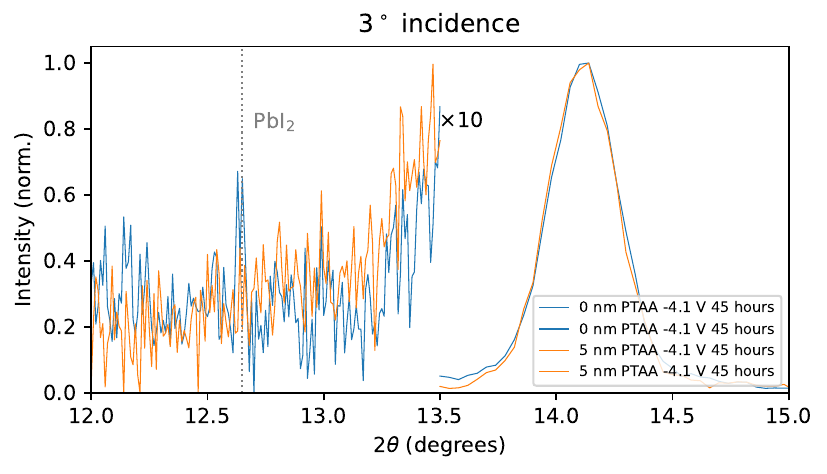}
	\caption{X-ray diffraction measurements at a fixed angle of incidence (3$^\circ$) of two perovskite p-i-n diodes after about 50 hours of fixed reverse bias ($-4.1$ V) stress. Blue: no HTL. Orange: 5-nm-thick PTAA HTL. The measurement was performed on the ``top" surface of the perovskite after delaminating the top metal contact on the ETL side of the cell. The incidence angle of 3$^\circ$ was chosen to probe most of the film thickness. The scattered signal is shown around the fundamental perovskite peak ($\approx 14.1^\circ$) and the expected location of the PbI$_2$ (001) peak ($\approx 12.65^\circ$). A very small, but not statistically significant, PbI$_2$ signal may be observed only in the cells without PTAA.
    }
\label{Fig_XRD}
\end{figure}

\begin{figure}[h!]
    \centering
    \includegraphics[width=5in]{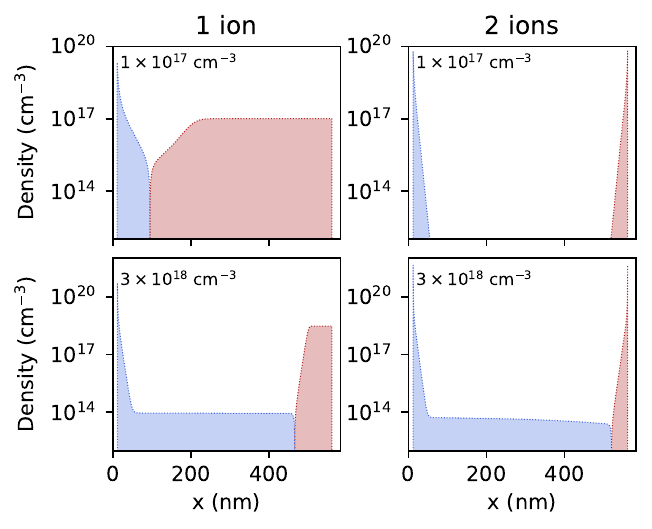}
	\caption{
    Net ionic charge density profiles calculated from drift-diffusion calculations. Left column: mobile cations and fixed anions. Right column: mobile cations and mobile anions. Top row: cations and anions in equal concentrations of $1\times10^{17}$ cm$^{-3}$. Bottom row: cations and anions in equal concentrations of $3\times10^{18}$ cm$^{-3}$. Blue shaded regions: positive charge. Red shaded regions: negative charge.
    }
\label{Fig_Mobile_Ion_Distributions}
\end{figure}

\begin{figure}[h!]
    \centering
    \includegraphics[width=7in]{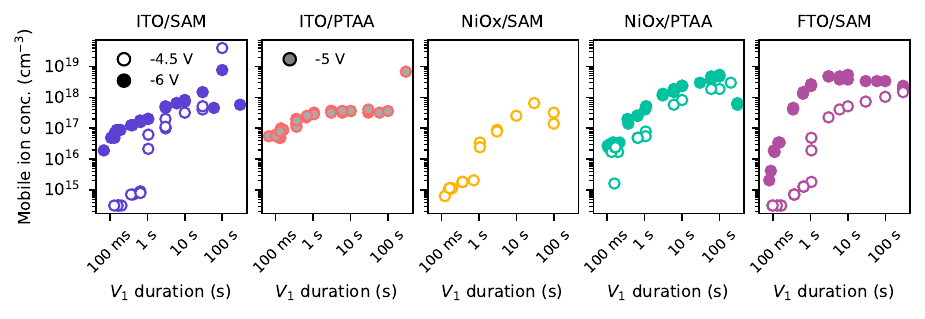}
	\caption{Estimated mobile ion concentrations versus $V_1$ hold duration for micro-cell (approximately 200 $\mu$m diameter) fabricated on a variety of TCO and HTL combinationvs: ITO/SAM, ITO/PTAA, NiO$_x$/SAM, NiO$_x$/PTAA and FTO/SAM (specified above each panel). The applied potential, $V_1$, is specified by the marker type (see legend). The perovskite composition in these cells is: FA$_{0.9}$MA$_{0.05}$Cs$_{0.05}$PbI$_3$ (bandgap approximately 1.54 to 1.55 eV). The NiO$_x$ film is a nanoparticle layer with thickness of about 20 nm. The ETL comprises 25 nm evaporated C$_{60}$ and 25 nm ALD SnO$_x$. The decreasing mobile ion concentrations in NiO$_x$/PTAA and FTO/SAM samples at long durations is possibly a measurement artifact seemingly related to very large reverse current densities flowing through the system before switching to 0 V.
    }
\label{Integrated_Current_Different_HTLs}
\end{figure}


\section{A survey of mobile ion measurement techniques and results}
Mobile ion concentrations have been inferred from several measurement techniques. Note that all experimental techniques exploit some model that describes the distribution of charges. Bertoluzzi pointed out that estimates may vary by several orders of magnitude depending on the model chosen to analyze the experimental results. Theoretical estimates have also been provided based on thermodynamics and activation energies of certain defects. \\

Current-transient measurements, sometimes referred to as bias-assisted charge extraction (BACE) measurements, have been used in references \cite{bertoluzzi_situ_2018, bertoluzzi_mobile_2020, penukula_quantifying_2023, thiesbrummel_ion-induced_2024, schmidt_characterization_2025}. All of these reports used ``preparation" voltages ($V_1$ in our notation) of about $+0.8$ V to $+1.2$ V. Ref. \cite{bertoluzzi_situ_2018} reported a mobile ion concentration of $\sim10^{18}$ cm$^{-3}$ in a formamidinium/Cs mixed halide perovskite. Ref. \cite{bertoluzzi_mobile_2020} reported mobile ion concentrations varying between $7\times10^{16}$ cm$^{-3}$ to $5\times10^{17}$ cm$^{-3}$ depending on the perovskite composition. The highest concentrations were estimated for compounds containing both methylammonium and bromide. The lowest concentrations were estimated for compounds containing formamidinium/Cs and iodide/bromide. Ref. \cite{penukula_quantifying_2023} reported values of $\sim10^{17}$ cm$^{-3}$ in  methylammonium lead iodide and $\sim10^{16}$ cm$^{-3}$ in a triple-halide composition. Ref. \cite{thiesbrummel_ion-induced_2024} reported mobile ion concentrations of $3.5\times10^{17}$ cm$^{-3}$ in a 1.63 eV triple-cation composition [Cs$_{0.05}$(FA$_{0.83}$MA$_{0.17}$)$_{0.95}$Pb(I$_{0.83}$Br$_{0.17}$)$_{3}$] before aging and final concentrations of about $4\times10^{18}$ cm$^{-3}$ after 10 hours of illumination at open-circuit conditions. Ref. \cite{schmidt_characterization_2025} reports approximately $2-3\times10^{17}$ cm$^{-3}$ in a Cs$_{0.05}$(MA$_{0.05}$FA$_{0.95}$)$_{0.95}$Pb(I$_{0.95}$Br$_{0.05}$)$_{3}$ composition. \\

Low-frequency capacitance measurements were used in ref. \cite{almora_capacitive_2015}. They estimated $2.4\times10^{17}$ cm$^{-3}$ for planar cells and $1.4\times10^{18}$ cm$^{-3}$ for mesoporous cells in methylammonium lead iodide perovskites. \\

Kelvin probe force microscopy was used in refs. \cite{weber_how_2018, birkhold_direct_2018}. Both of these works estimated ion concentrations of about $2\times10^{15}$ cm$^{-3}$ based on the observed band bending and an ion depletion model comparable to models used in the references above. In ref. \cite{weber_how_2018}, a formamidinium/methylammonium mixed-halide lead perovskite composition was used. Samples were cleaved using a focused ion beam in order for cross-sections to be measured. They mentioned that their estimated ion concentration is orders of magnitude lower than estimates of $\sim 10^{19}$ cm$^{-3}$ based on the lowest activation energy of about 0.6 eV for I$^-$ migration through vacancies \cite{eames_ionic_2015}. However, in a different work \cite{walsh_self-regulation_2015}, reaction energies of 0.22 eV were associated with equilibrium concentrations of about $10^{18}$ cm$^{-3}$, implying that 0.6 eV may be associated with much lower concentrations of $6\times10^{16}$ cm$^{-3}$. Measuring the cross-section of a cleaved sample may alter charge dynamics in the cell which will not necessarily represent the behavior in the bulk. In ref. \cite{birkhold_direct_2018}, lateral devices comprising a methylammonium lead iodide perovskite were measured and no cleaving was necessary.

\section{Drift-diffusion band-diagram calculations}
We used \emph{Driftfusion}, an open source drift-diffusion code based in \emph{MATLAB} for simulating semiconductor devices with mixed ionic-electronic conducting materials \cite{calado_driftfusion_2022}. The material and device input files and analysis scripts used to calculate energy diagrams are included as Supplemental Files with this publication. Calculation methods generally used the `doCV()' method to perform cyclic-voltammetry (CV) sweeps of the applied potential. We explored ion mobilities and CV sweep rates to ensure ionic equilibrium was reached. Energy diagrams shown in the manuscript represent the equilibrium system. For those calculations, transport layers were assumed to be highly doped so that the potential drop across the transport layers is negligible. \\

To assess whether mobile ions are expected to fully equilibrate during our fastest reverse-JV scans (about 10 V/s), we performed CV-type calculations from 0 V to -4 V at a constant sweep rate of 10 V/s while varying the ion mobility between $10^{-10}$ cm$^2$/Vs to $10^{-7}$ cm$^2$/Vs. The upper limit of mobility ($10^{-7}$ cm$^2$/Vs) is comparable to the estimate from Ref. \cite{bertoluzzi_situ_2018}. A lower commonly used value is $\approx 3\times10^{-9}$ cm$^2$/Vs \cite{schmidt_characterization_2025}. For these calculations, we kept the mobile cation concentration constant at $3\times10^{17}$ cm$^{-3}$. Figure \ref{Fig_SI_Drift_Diffusion_Calcs} shows energy diagrams calculated at -4 V during the CV sweep from 0 V to $-4$ V at 10 V/s for varying mobile cation concentrations. Equilibrium is assessed by inspecting the slope of the CBM and VBM outside the ionic depletion region (toward the ``left" side of the film in the calculations). When ions are fully equilibrated, the VBM and CBM become flat in this region. For scan rates up to 10 V/s, ionic equilibrium is evidently a reasonable assumption for cation mobilities exceeding about $3\times10^{-9}$ cm$^2$/Vs. The most important typical input parameters are shown in Tables 1 and 2.

\begin{table}[h]
\centering
\caption{Important \emph{Driftfusion} input parameters (part 1)}
\small
\begin{tabular}{l l c c c c c}
\hline
Layer & Material & Thickness (cm) & EA (eV) & IP (eV) & EF0 (eV) & Et (eV) \\
\hline

Electrode & -- & -- & -- & -- & -5.49 & -- \\

Layer & PTAA & $1.0\times10^{-6}$ & -1.8 & -5.5 & -5.49 & -4.05 \\

Active & MAPICl & $5.5\times10^{-5}$ & -3.9 & -5.5 & -5.2 & -4.6 \\

Layer & C60 & $2.0\times10^{-6}$ & -3.9 & -6.2 & -3.91 & -5.0 \\

Electrode & -- & -- & -- & -- & -3.91 & -- \\

\hline
\end{tabular}
\end{table}

\begin{table}[h]
\centering
\caption{Important \emph{Driftfusion} input parameters (part 2)}
\small
\begin{tabular}{l c c c c c c c}
\hline
Material & $N_c/N_v$ (cm$^{-3}$) & $\mu_n/\mu_p$ (cm$^2$ V$^{-1}$ s$^{-1}$) & $N_{\mathrm{ion}}$ (cm$^{-3}$) & $\mu_{\mathrm{cation}}$/$\mu_{\mathrm{anion}}$ (cm$^2$ V$^{-1}$ s$^{-1}$) & $\varepsilon_r$ & $\tau_n/\tau_p$ (s) & $s_n/s_p$ (cm s$^{-1}$) \\
\hline
Electrode & -- & -- & -- & -- & -- & -- &$10^{7}/10^{7}$ \\

PTAA & $10^{21}/10^{21}$ & $0.1/0.1$ & $3\times10^{17}$ & 0/0 & 4.0 & $10^{-6}/10^{-6}$ & -- \\

Interface & $10^{19}/10^{19}$ & $10/10$ & $3\times10^{17}$ & 0/0 & 23.0 & $10^{-7}/10^{-7}$ & $10^3/10^7$ \\

MAPICl & $10^{19}/10^{19}$ & $20/20$ & $3\times10^{17}$ & $10^{-7}/0$ & 23.0 & $10^{-7}/10^{-7}$ & -- \\

Interface & $10^{19}/10^{19}$ & $10/10$ & $3\times10^{17}$ & 0/0 & 23.0 & $10^{-7}/10^{-7}$ & $10^7/10^3$ \\

C60 & $10^{21}/10^{21}$ & $0.1/0.1$ & $3\times10^{17}$ & 0/0 & 3.0 & $10^{-6}/10^{-6}$ & -- \\

Electrode & -- & -- & -- & -- & -- & -- &$10^{7}/10^{7}$ \\

\hline
\end{tabular}
\end{table}


\section{Tunneling current calculations}
Tunnel currents in our system were calculated from the model described by Chang and Sze \cite{chang_carrier_1970}. Their model was developed for a metal-semiconductor junction, i.e., a Schottky junction. They included both thermionic and tunneling terms in their treatment. In our system, tunneling happens at the interface between the perovskite and the electron transport layer (ETL). We thus adapted Chang and Sze's treatment to the current scenario in which a perovskite (semiconductor) is in direct contact with an ETL. We first approximate the ETL as a metal. We then change the metal density of states to represent the scenario with an ETL, which will be approximated as an n-type doped semiconductor. Thermionic emission is neglected because the barrier is too large for that term to be relevant at room temperature. \\

\begin{figure}[h!]
    \centering
    \includegraphics[width=4in]{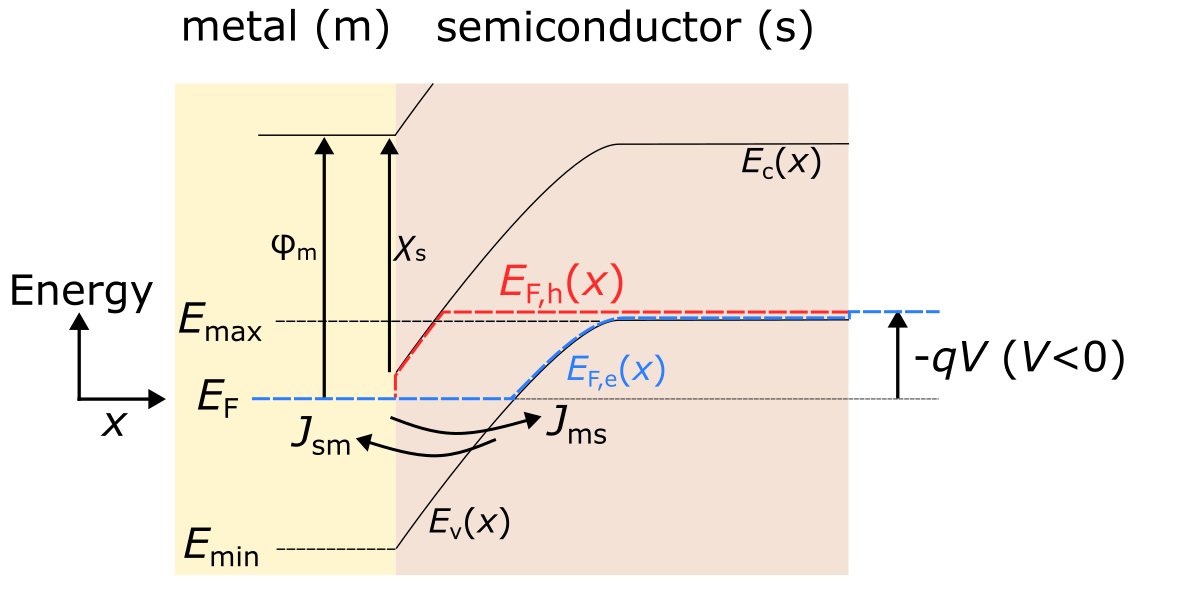}
	\caption{
        Energy diagram to calculate tunneling currents in our system, which is treated as a metal-semiconductor junction. All energy levels necessary for performing the calculation are specified. 
        }
\label{SIfig:Sze_Tunneling}
\end{figure}

The energy structure of the system is described in Fig. \ref{SIfig:Sze_Tunneling}. A metal (m) and a semiconductor (s) are in direct contact. The horizontal axis represents the growth axis of the material system ($x$). We define $x=0$ as the junction position with the semiconductor in the region $x>0$. The Fermi level in the mtal is $E_F$, which we define as $E_F=0$ (i.e. all energies are referenced with respect to the Fermi level in the metal). The metal's work function $\phi_m$ is defined as the energy difference between $E_F$ and vacuum. The semiconductor's electron affinity $\chi_s$ is defined as the energy difference between the semiconductor's conduction band minimum (CBM) and vacuum. The energy alignment between the metal and semiconductor is defined by $\chi_x - \phi_m$. The CBM energy $E_c(x)$ depends on $x$. The semiconductor's valence band maximum (VBM) energy $E_v(x)$ also depends on $x$. The CBM and VBM are related everywhere in the semiconductor by the bandgap energy $E_g$ by $E_c(x) = E_v(x) + E_g$.The electron Fermi level $E_{F,e}(x)$ (blue dashed curve) and hole Fermi level $E_{F,h}(x)$ are generally $x$ dependent. We use the results from Bertoluzzi \emph{et. al.} \cite{bertoluzzi_mobile_2020} to define the electron and hole Fermi levels. Under an applied bias $V$, they will not be equal. For our calculations, we assume $E_{F,e}(x) = E_v(x)$ where $E_v(x)\ge0$. We assume $E_{F,e}(x)=0$ where $E_v(x)<0$. Similarly, we assume $E_{F,h}(x) = -qV$ where $E_c(x)>-qV$. We assume $E_{F,eh}(x)=E_x(x)$ where $E_c(x)<-qV$. These energy levels are specified in Fig. \ref{SIfig:Sze_Tunneling}. \\

The total tunneling current contains a contribution due to tunneling from metal to semiconductor with current density $J_{ms}$ and from semiconductor to metal with current density $J_{sm}$. It takes into account the density of states and the Fermi occupancies of electrons and holes in the metal and in the semiconductor. All tunneling happens to/from the semiconductor's valence band. The energy-dependent Fermi occupancy of electrons in the metal is:

\ba{
F_{m,e}(E) = \frac1{1+\exp\left(\frac{E - E_{F}}{k_B T}\right)}
}

The Fermi occupancy of electrons in the semiconductor depends on $x$ and energy:

\ba{
F_{s,e}(x,E) = \frac1{1+\exp\left(\frac{E - E_{F,e}(x)}{k_B T}\right)}
}

The Fermi occupancy of holes in the metal is:

\ba{
F_{m,h}(E) = \frac1{1+\exp\left(\frac{-E + E_{F}}{k_B T}\right)}
}

The Fermi occupancy of holes in the semiconductor is:

\ba{
F_{s,h}(x,E) = \frac1{1+\exp\left(\frac{-E + E_{F,h}(x)}{k_B T}\right)}
}

The total current density is $J = -J_{ms} + J_{sm}$. It has an implicit dependence on the applied potential because the Fermi occupancies depend on the applied potential. Each term is calculated essentially by integrating over all energies the product of three terms: 1) The volume density of charges at that energy/position, 2) the thermal velocity, and 3) a tunneling probability. The thermal velocity is captured by the so-called effective Richardson constant (or Richardson velocity) of the semiconductor $A^*$. This Richardson constant will be discussed in more detail later. The tunneling probability $T(E)$ depends on energy. It depends on the electron's or hole's energy with respect to the tunneling barrier ``height" which is $-E_v(x)$. We assume the tunneling probability is equal for electrons and holes. In reality, it has a square-root dependence on the carrier's effective mass.  \\

We write $J_{ms}$ and $J_{sm}$ separately:

\ba{
J_{ms} 
&= - \frac{A^* T}{k_B} \int_{E_{min}}^{E_{max}} 
dE \ T(E) \{F_{m,e}(E)[1-F_{s,e}(E)] + F_{m,h}(E)[1-F_{s,h}(E)]\} \\
J_{sm} 
&= \frac{A^* T}{k_B} \int_{E_{min}}^{E_{max}} 
dE \ T(E) \{F_{s,e}(E)[1-F_{m,e}(E)] + F_{s,h}(E)[1-F_{m,h}(E)]\}
}
 
\noindent where $k_B$ is the Boltzmann constant and $T$ is the temperature. The tunneling probability is defined by integrating the imaginary part of the charge's wavenumber through the barrier:

\ba{
T(E) = \exp \left\{ 
-\frac{2}{\hbar} \int_0^{x_{max}} |\sqrt{2m[E_v(x) - E]}| dx 
\right\}
}

\noindent where $x_{max}$ is the $x$ position where $E_v(x)$ crosses $E$, i.e., defined by $E_v(x_{max})=E$. Note that $E_v(x)<E$ in the tunneling region $0<x<x_{max}$, and so the value under the square root is negative and the square root evaluates to a purely imaginary value. The vertical bars indicate we take the absolute value, ultimately leading to the tunneling term being a decaying exponential in $x$. \\

For the effective Richardson constant $A^*$, we use the definition described by Chang and Sze \cite{chang_carrier_1970}: $A^* = g m^* A$, where $m^*$ is the carrier effective mass in the relevant band (here, the VBM), $g$ is the band degeneracy, and $A$ is the free-space Richardson constant. The free-space Richardson constant is taken to be $A$=120 A/cm$^2$/K$^2$, which can be derived from the expression $A=4\pi m_0 q k^2/h^3$ by taking $k$ to be the free-space wave number of an electron with thermal energy at 300 K ($m_0$ is the free-space electron mass, $q$ the elementary charge, $h$ is Planck's constant). Here, we take $m^*=0.25 m_0$ for the VBM. The degeneracy of the VBM is 1. (This contrasts with silicon or III-V semiconductors in which the degeneracy of the VBM is 3 and the degeneracy of the CBM is 1.)  The effective Richardson constant in our system is therefore $A^* = g m^* A \approx 30$ A/cm$^2$/K$^2$. \\

For the energy levels, we used the values outlined in Table 3.\\

\begin{table}[h!]
\centering
\caption{Tunneling calculation parameters}
\begin{tabular}{|c|c|}
\hline
Parameter & Value used \\ \hline
Built in voltage & 0.8 V \\ \hline
Metal work function & 4.6 eV \\ \hline
Semiconductor electron affinity & 3.91 eV \\ \hline
Semiconductor relative permittivity & 25 \\ \hline
Relative effective mass & 0.3 \\ \hline
Temperature & 300 K \\ \hline
Band Gap & 1.65 eV \\ \hline
\end{tabular}
\end{table}

We note a few limitations of this model. 1) It ignores the ETL. The most precise treatment would include charge densities and band bending through the ETL. The electron and hole density of states in the ETL, however, are difficult to know or estimate. Our treatment may be reasonable if charges can tunnel into the VBM of the ETL, or if there is a high enough concentration of trap states to mediate the tunneling into the perovskite. 2) It does not account for image-charge effects which would reduce the barrier height and \emph{increase} the tunneling rate \cite{chang_carrier_1970}. 3) The tunneling expression (Eqn. 7) ignores the possibility for internal wave-function reflections in the barrier. Reflections may increase or decrease the transmittance depending on a wide range of variables. 

\section{Order-of-magnitude estimate of mobile ion concentrations under oxidizing conditions}
We provide an order-of-magnitude estimate for the average volume density of mobile vacancies that may be generated under highly oxidizing conditions such as reverse bias. The concept is illustrated in Fig. \ref{SI_Layer_of_Oxidation}. Assume iodide become oxidized in each unit cell at an interface, just one single unit cell deep (left panel illustration). The resulting iodine is represented by orange circles. Each oxidized iodide dissociates from the lattice, leaving a new iodine vacancy. Unit cells with an iodine vacancy are represented by a yellow rotated square. The \emph{areal} density of unit cells with a new iodine vacancy is calculated to be about $1.25\times10^{14}$ cm$^{-3}$, assuming the unit cell is square with length 9 \AA. This areal density corresponds to an \emph{average volume concentration} of about $2.5\times10^{18}$ cm$^{-3}$ (right panel illustration). For example, given enough time, diffusion in the absence of a built-in potential would uniformly distribute these vacancies. \\

\section{Estimating mobile ion concentrations using current-transient electrical measurements}

Here we describe how to relate external currents to internal ion motion while accounting for potential variation across the transport layers. Measurement methods are described in the experimental methods section in this document. \\

By definition, the potential $V$ across the junction is the integral of the electric field $E(x,t)$ across the junction. In this problem, the total electric field in the cell has electronic contributions (from the electrodes) and ionic contributions (from the perovskite). The charge distributions and the resulting fields generally depend on position ($x$) and time ($t$). Let the perovskite exist only within $0<x<d_\text{pvsk}$. Let the ETL exist within $d_\text{pvsk}<x<d_\text{pvsk}+d_\text{ETL}$. For convenience we define $d \equiv d_\text{pvsk}+d_\text{ETL}$. The ionic charge is distributed across the perovskite layer, described by a volume density $\rho_\text{ion}(x,t)$ for $0<x<d_\text{pvsk}$. Assume the ETL  has zero charge density ($\rho(x)=0$ for $d_\text{pvsk}<x<d$). The HTL will be assumed to be thin enough that the charge and potential drop across the HTL can be ignored in this problem. Electrode boundaries exist at $x=0$ and $x=d$. Let the electronic charge areal density on the electrodes be called $q(t)$; it is concentrated completely at the metal-dielectric interfaces and will be treated as a two-dimensional charge distribution ($\rho_e(x) = -q\delta(x) + q(x-d)$). We will use a constant high-frequency permittivity $\epsilon_\text{pvsk} \approx 25$ to account for fast-responding electronic permittivity contribution in the perovskite which \emph{does not} depend on this ``slow" ionic motion. (I.e., ions will be treated as free charges on top of the background dielectric response.) For the ETL, assume a permittivity $\epsilon_\text{ETL} \approx 4$. A significant portion ($\approx$25-30\%) of the applied potential is dropped across the ETL under reverse bias because it has a low doping and a low permittivity. We will quantify that portion later in this section. The effect of charge accumulation and depletion in transport layers was previously described in Ref. \cite{schmidt_characterization_2025}. \\

The general charge distribution is summarized as follows:

\ba{
\rho(x,t) &= -q(t)\delta(x) + q(t)\delta(x-d) \hspace{1cm} \text{(electrodes at $x=0$ and $x=d$)} \\
\rho(x,t) &= \rho_\text{ion}(x,t) \hspace{3.5cm} 0<x<d_\text{pvsk} \text{  (ions in perovskite)} \\
\rho(x,t) &= 0 \hspace{4.6cm} d_\text{pvsk} < x < d \text{  (ETL)}
}

We start with the general expression for the total displacement field in the cell ($0<x<d$),

\ba{
D(x,t) 
= \int_0^x \rho(x',t) dx' 
= -q(t) + \int_0^x \rho_\text{ion}(x',t) dx'
}

\noindent and $D=0$ outside that region. The electric field at position $x$ in the cell, $E(x)$, is related to $D(x)$ by the local permittivity $\epsilon(x)$:

\ba{
E(x,t) 
= \frac{D(x,t)}{\epsilon(x)}
= -\frac{q(t)}{\epsilon(x)} 
+ \frac1{\epsilon(x)} \int_0^x \rho_\text{ion}(x',t) dx'
}

\noindent and $E=0$ outside of that region. The potential across the junction is

\ba{
V(t) 
&= - \int_0^{d} E(x,t) dx \\
&= q(t) 
\left(
\frac{d_\text{pvsk}}{\epsilon_\text{pvsk}}
+ \frac{d_\text{ETL}}{\epsilon_\text{ETL}}
\right)
- \int_0^d \frac1{\epsilon(x')}
\left[
\int_0^{x'} \rho_\text{ion}(x'',t) dx'' 
\right] dx' \\
&= q(t) 
\left(
\frac{d_\text{pvsk}}{\epsilon_\text{pvsk}}
+ \frac{d_\text{ETL}}{\epsilon_\text{ETL}}
\right)
- \frac1{\epsilon_\text{pvsk}} \int_0^{d_\text{pvsk}} 
\left[
\int_0^{x'} \rho_\text{ion}(x'',t) dx'' 
\right] dx' 
- \frac1{\epsilon_\text{ETL}} \int_{d_\text{pvsk}}^{d_\text{ETL}} 
\left[
\int_0^{x'} \rho_\text{ion}(x'',t) dx'' 
\right] dx' 
}

In the final integral in the line above, $\int_0^{x'}\rho_\text{ion}(x'',t)dx''$ is a constant for $x'>d_\text{pvsk}$. We can also exploit the fact that $\int_0^{d_\text{pvsk}}\rho_\text{ion}(x'',t)dx'' = 0$ (the perovskite is net charge neutral since ions only redistribute), and so we can rewrite the last expression:

\ba{
V(t) &= 
q(t) 
\left(
\frac{d_\text{pvsk}}{\epsilon_\text{pvsk}}
+ \frac{d_\text{ETL}}{\epsilon_\text{ETL}}
\right)
- \frac1{\epsilon_\text{pvsk}} \int_0^{d_\text{pvsk}} 
\left[
\int_0^{x'} \rho_\text{ion}(x'',t) dx'' 
\right] dx' 
- \frac{d_\text{ETL}}{\epsilon_\text{ETL}} 
\int_0^{d_\text{pvsk}} \rho_\text{ion}(x'',t) dx''  \\
&= q(t) 
\left(
\frac{d_\text{pvsk}}{\epsilon_\text{pvsk}}
+ \frac{d_\text{ETL}}{\epsilon_\text{ETL}}
\right)
- \frac1{\epsilon_\text{pvsk}} \int_0^{d_\text{pvsk}} 
\left[
\int_0^{x'} \rho_\text{ion}(x'',t) dx'' 
\right] dx'
}

In the treatment above, we have assumed a very thin HTL is used and so the potential drop across that layer is ignored. In a cell with a thicker HTL, a third linear term $q(t)d_\text{HTL}/\epsilon_\text{HTL}$ will need to be added to Eqn. 17. If the permittivity of the HTL is comparable to that of the ETL, then the total potential drop across both transport layers can be approximated by considering the total transport layer (TL) thickness: $d_\text{ETL}/\epsilon_\text{ETL} + d_\text{HTL}/\epsilon_\text{HTL} \approx (d_\text{ETL}+d_\text{HTL})/\epsilon_\text{TL}$. 

Eqn. 17 gives a relationship between the electronic charge on the electrodes and the ionic charge distribution in the perovskite at any time $t$:

\ba{
q(t) = 
\frac{V(t)}
{ \frac{d_\text{pvsk}}{\epsilon_\text{pvsk}}
+ \frac{d_\text{ETL}}{\epsilon_\text{ETL}} } + 
\frac1 {d_\text{pvsk} + d_\text{ETL} \frac{\epsilon_\text{pvsk}}{\epsilon_\text{ETL}}}
\int_0^{d_\text{pvsk}} 
\left[
\int_0^{x'} \rho_\text{ion}(x'',t) dx'' 
\right] dx' \\
}

This expression is easy to understand. The first term is the charge expected from the parallel-plate capacitor holding a voltage $V(t)$ and an effective dielectric constant of $\frac1d (d_\text{pvsk}\epsilon_\text{pvsk} + d_\text{ETL}\epsilon_\text{ETL})$. The second term ``removes" electrons because the ions maintain some portion of the applied potential. To gain some intuition for this expression, momentarily assume the ETL is absent ($d_\text{ETL}=0$) and all of the ionic charge is distributed in a 2D layer at each interface: $\rho(x,t) = \sigma(t)\delta(x) - \sigma(t)\delta(x-d_\text{pvsk})$. We get 

\ba{
q(t) = \frac{\epsilon_\text{pvsk}}{d_\text{pvsk}} V(t) + \sigma(t)
}

\noindent which just says that $q(t)$ and $\sigma(t)$ together sustain the potential $V(t)$ across the cell. If $V(t)$ does not change in time ($dV(t)/dt=0$) while charges are allowed to move, the electronic and ionic \emph{currents} are identical: $\partial q(t)/\partial t = \partial \sigma(t)/\partial t$. When the ions are distributed differently, the ionic and electronic currents differ. \\

Ultimately, we do not care about specific dynamics because we will be integrating over time immediately after a change in potential until all charges have stopped moving. This amounts to evaluating the current difference: $\Delta q = q(t\rightarrow\infty) - q(0)$ where $t=0$ marks the moment the potential across the junction was switched. We will make the assumption that the potential across the junction for $t\geq0$ is the instantaneously applied potential $V$. The charge difference is

\ba{
\Delta q 
&= q(t\rightarrow\infty) - q(0) \\
&= \frac1 {d_\text{pvsk} + d_\text{ETL} \frac{\epsilon_\text{pvsk}}{\epsilon_\text{ETL}}}
\int_0^{d_\text{pvsk}} \left\{\int_0^{x} 
\left[\rho_\text{ion}(x', t\rightarrow\infty) - \rho_\text{ion}(x', 0) \right] dx' \right\}dx
}

It is clear from this expression that ions redistributing inside the cell $[\rho(x', t\rightarrow\infty) - \rho(x', 0) \neq 0]$ lead to a change in the electronic charge on the electrodes. \\

To complete the calculation, we now need a model for the function $\rho(x,t)$. For potentials $V\leq0$, we will use the single mobile-ion (cation) model from Bertoluzzi \emph{et al.} \cite{bertoluzzi_mobile_2020}. It consists of a positively charged Debye layer of width $L_D=\sqrt{\epsilon_\infty V_T / (2eN_0)}$ at the HTL side of the perovskite ($0 < x < L_D$) and a negatively charged vacancy depletion region of width $w_-$ at the ETL side of the perovskite ($d_\text{pvsk}-w_- < x < d_\text{pvsk}$). The charge density in the depletion region is assumed to be the mobile ion density $-eN_0$. The charge density in the Debye layer, $en_+$ is derived by assuming charge neutrality between the depletion and accumulation regions: $en_+ L_D=eN_0 w_-$. The depletion width $w_-(V)$ depends on the voltage held across the perovskite layer. The accumulation region density $n_+$ is thus implicitly voltage dependent. Nothing else in the system is voltage dependent. We will thus give $\rho_\text{ion}(x,t)$ a dependence on two voltages: $V_1$ will represent the voltage at which the ionic charge distribution was previously equilibrated \emph{before the voltage was rapidly switched} at time $t=0$. It will describe the charge distribution at time $t\leq0$. $V_2$ will be the voltage \emph{immediately after switching}. It will describe the charge distribution at times that are sufficiently long ($t\rightarrow\infty$) for ions to equilibrate to $V_2$. $V_1$ and $V_2$ are the potentials maintained across the perovskite layer; they will both be smaller in magnitude than the applied potential. \\

We write the ionic charge densities explicitly:

\ba{
\rho(V_1, V_2; x,t\rightarrow\infty) &= 
\begin{cases}
\frac{e N_0 w_-(V_2)}{L_D} & 0<x<L_D \\
0 & L_D < x < d_\text{pvsk}-w_-(V_2) \\
-e N_0 & d_\text{pvsk}-w_-(V_2) < x < d_\text{pvsk}
\end{cases} \\
\rho(V_1, V_2; x, 0) &= 
\begin{cases}
\frac{e N_0 w_-(V_1)}{L_D} & 0<x<L_D \\
0 & L_D < x < d_\text{pvsk}-w_-(V_1) \\
-e N_0 & d_\text{pvsk}-w_-(V_1) < x < d_\text{pvsk}
\end{cases}
}

We make one final simplification to the expression above based on the assumption that the ionic charge distribution is allowed to fully equilibrate between voltages $V_1$ and $V_2$. This allows us to eliminate the time variable.

\ba{
\Delta q (V_1, V_2)
= \frac1 {d_\text{pvsk} + d_\text{ETL} \frac{\epsilon_\text{pvsk}}{\epsilon_\text{ETL}}}
\int_0^{d_\text{pvsk}} \left\{\int_0^{x} 
\left[\rho_\text{ion}(V_2; x') - \rho_\text{ion}(V_1; x') \right] dx' \right\}dx
}

\noindent where $\rho(V; x)$ is simply the fully equilibrated ionic charge density at potential $V$. \\

For $V_1=+0.8$ V, we assume a nearly flat-band condition with $\rho(x)=0$. \\

We next need to relate the voltage held across the perovskite ($V_1$ and $V_2$) to the applied potential ($V$). Let the applied potential ($V$) be the sum of the potentials across the perovskite ($V_\text{pvsk}$) and ETL ($V_\text{ETL}$): $V=V_\text{pvsk} + V_\text{ETL}$. Based on the dielectric model above, the displacement field in the ETL originates completely from the charge on the electrodes (no internal charge) and is a constant $D$. We assume a ``linear" approximation and let the displacement field also be a constant in the perovskite layer (i.e., no charge in perovskite). The potential drop across each layer then depends only on the relative permittivities and thicknesses of each layer: $V_\text{pvsk} = D d_\text{pvsk}/\epsilon_\text{pvsk}$ and $V_\text{ETL} = D d_\text{ETL}/\epsilon_\text{ETL}$. Then,

\ba{
V
= V_\text{pvsk}+V_\text{ETL} 
= V_\text{pvsk} \left( 1 + \frac{V_\text{ETL}}{V_\text{pvsk}} \right)
= V_\text{pvsk} (1 + \frac{d_\text{ETL} \epsilon_\text{pvsk}}{d_\text{pvsk} \epsilon_\text{ETL}}) 
\longrightarrow 
V_\text{pvsk} 
= \frac{V}{(1 + \frac{d_\text{ETL} \epsilon_\text{pvsk}}{d_\text{pvsk} \epsilon_\text{ETL}})}
}

\begin{figure}[h!]
    \centering
    \includegraphics[width=3.5in]{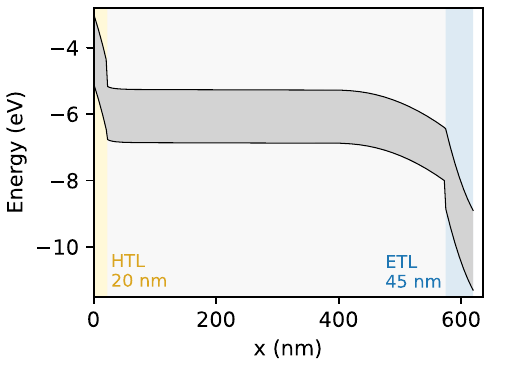}
	\caption{
        Drift-diffusion band diagram calculation with mobile cations ($N_\text{ion} = 1\times10^{17}$ cm$^{-3}$). The HTL (10 nm thick) and ETL (45 nm thick) are assumed to have low doping ($\sim10^{15}$ cm$^{-3}$) and relative permittivity $\epsilon=4$. The perovskite has $\epsilon=23$. The applied potential is $-4$ V.
        }
\label{SIfig:Drift_Diffusion_Band_Diagram_w_HTL_ETL}
\end{figure}

Taking values of $d_\text{ETL}=60$ nm, $\epsilon_\text{ETL}=4$, $d_\text{pvsk}=600$ nm and $\epsilon_\text{pvsk}=25$, the voltage drop across the perovskite is about 60-65\% of the applied potential. As specified just below Eqn. 17, the potential drop across both transport layers can be approximated by summing the total transport layer thickness as long as the permittivity is taken to be the average of those layers. This effect can be seen in rigorous drift-diffusion calculations of perovskite diode band diagrams, as shown in previous reports \cite{schmidt_impact_2023} and in our calculations (Fig. \ref{SIfig:Drift_Diffusion_Band_Diagram_w_HTL_ETL}). \\

The calculated displaced electronic charge vs. $N_0$ is plotted in Fig. \ref{SIfig:Displaced_Charge_vs_N0} for a range of applied potentials $V_1$, including $V_1=+0.8$ V (left panel) and reverse potentials $V_1<0$ V (right panel). All calculations assume $V_2=0$ and the single mobile ion (cation) depletion approximation model described above. We calculate one set of curves assuming no ETL (the applied potential is completely dropped across the perovskite layer; solid curves), which would represent a case with highly doped contact layers. We calculate a second set of curves using the model describes above with a 60-nm thick ETL (30 nm C$_{60}$ plus 30 nm SnO$_x$), with $\epsilon_\text{pvsk}=25$ and $\epsilon_\text{ETL}=4$ (dashed curves). The perovskite film is 600 nm in all cases. A ``kink" in the curves around $\approx10^{16}$cm$^{-3}$ arises from calculation artifacts when the ionic depletion width reaches the perovskite film thickness. The ETL-corrected expression requires about a 5-times higher ion concentration than the no-ETL treatment, primarily because of a non-negligible ($\approx$25-30\%) potential drop across the low-permittivity and relatively thick ($\approx$60 nm) ETL. The values specified in our manuscript are derived from the ETL-corrected treatment (dashed lines). \\

A summary of the assumptions used is listed in Table 4 with important device parameters in Table 5. \\

We would like to highlight our assumption of a one-mobile-ion distribution model proposed by Bertoluzzi \emph{et al.} \cite{bertoluzzi_mobile_2020}. We used this assumption to create an analytical solution to the electrostatics problem to relate external displaced charge to mobile ion concentration. While other methods, such as drift-diffusion simulations, can be used to relate the measured integrated current to the mobile ion concentration using one or two mobile ions, rigorous comparison to experiments is challenging. Values up to $10^{18}$ cm$^{-3}$ and $4\times10^{18}$ cm$^{-3}$ were inferred by M. Schmidt et al. \cite{schmidt_quantification_2026} and J. Thiesbrummel et al. \cite{thiesbrummel_ion-induced_2024} using these same techniques. The saturation of capacitance discussed in the main text causes saturation in drift-diffusion calculations at a lower displaced charge than in our measurements (also seen previously \cite{schmidt_quantification_2026}), so drift-diffusion simulations only provide a \emph{lower limit} for our inferred final mobile ion concentration (at 200 s in Fig. 3b). Preliminary drift-diffusion calculations suggest that adding a mobile anion to drift-diffusion simulations with a $\sim$1000$\times$ lower mobility than the cation decreases the \emph{lower limit} for the final mobile ion concentration by about 2$\times$, while an anion with $\sim$100$\times$ lower mobility yields a lower limit about 5$\times$ smaller.  \\

\begin{table}[h!]
\centering
\caption{Assumptions used for mobile ion concentration calculations}
\begin{tabular}{|c|c|}
\hline
\textbf{Assumptions} & \textbf{Description} \\
 \hline
ETL neutrality & The ETL is undoped (no charge density) \\
Thin HTL & The HTL is thin enough that there is no potential drop \\
2D electrode charge & Charge on electrodes is treated as a 2D distribution at the ETL-electrode interface \\
Free ionic charges & Ions are treated as free charges on top of the dielectric background \\
Perovskite neutrality & Equal numbers of cations and anions are assumed (overall charge neutral) \\
Instantaneous potential & The potential across the junction is the applied potential at all times $t \geq 0$ \\
Ionic distribution model & Ions are distributed according to Bertoluzzi \emph{et al.} \cite{bertoluzzi_mobile_2020} \\
Full equilibration & Ionic distribution fully equilibrates by the end of transient measurements \\
Flat-band condition at +0.8 V & Ions are mostly uniformly distributed at +0.8 V \\ \hline
\end{tabular}
\end{table}

Only when calculating the potential drop across the perovskite layer vs. the applied potential, we assume a linear potential variation across each layer (i.e., no charges in the perovskite).

\begin{table}[h!]
\centering
\caption{Important parameters used in mobile ion concentration calculations}
\begin{tabular}{|c|c|}
\hline
\textbf{Parameters} & \textbf{Values} \\
 \hline
 Perovskite relative permittivity & 25 \\
 Perovskite thickness & 600 nm \\
 ETL relative permittivity & 4 \\
 ETL thickness & 60 nm \\
\hline
\end{tabular}
\end{table}

\section{Analyses of ITO Texture and PTAA coverage}
\begin{figure}[h!]
    \centering
    \includegraphics[width=6in]{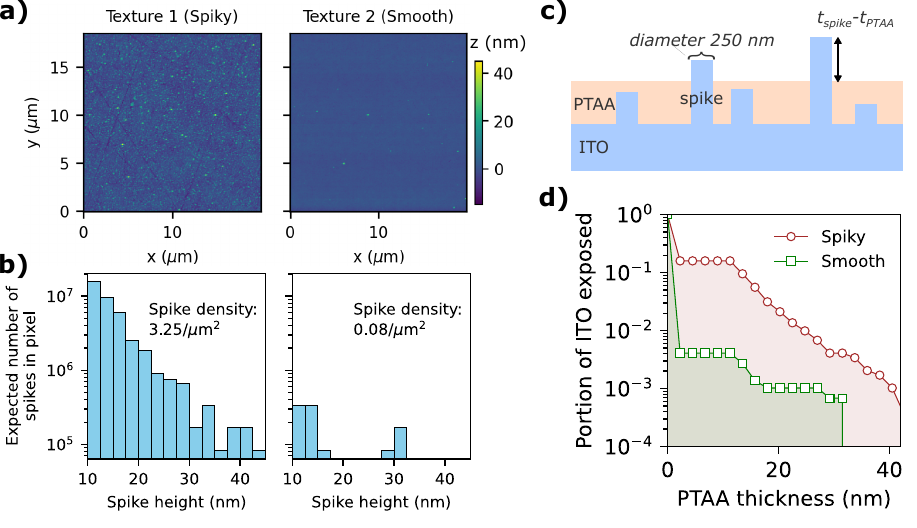}
	\caption{
    \textbf{(a)} AFM topography images of the two ITO substrate textures used in this study. 
    \textbf{(b)} The AFM images were processed to count individual ``spikes". Spike heights ($t_\text{spike}$) were recorded and the number of spikes with height $t_\text{spike}$ are histogrammed. The areal density of spikes is scaled to the size of a typical cell in our study (12 mm$^2$). 
    \textbf{(c)} Cross-sectional illustration of spikes on ITO with a PTAA HTL used to estimate the portion of exposed ITO (panel d). 
    \textbf{(d)} Estimated portion of ITO exposed (vertical axis) when a PTAA film with a nominal thickness (horizontal axis) is spin-cast on top of the ITO. For this calculation, spikes are treated as circular cylinders (rectangles in the cross-sectional illustration) with radius 125 nm (inferred from AFM images). The PTAA is treated as a fluid with zero viscosity. Calculations use the spike statistics shown in panel b. Estimates for ``spiky" (red) and ``smooth" ITO are shown. 
    }
\label{Fig_SI_ITO_Spikes_Analysis}
\end{figure}

Fig. \ref{Fig_SI_ITO_Spikes_Analysis}a shows atomic-force microscopy (AFM) topography images of two distinct ITO textures used in our work, originating from two different commercial manufacturers and two different cleaning procedures. One is highly textured with a typical minimum/maximum surface roughness of approximately $\pm$15 nm and a high concentration of spikes exceeding 10-40 nm (left image). We refer to this texture as ``Spiky". Note that the root-mean-square roughness of these surfaces is still only about 2 nm, highlighting the need for more rigorous ITO inspection before device fabrication. The second ITO texture is very smooth after vigorous mechanical brushing and cleaning, showing a small minimum/maximum surface roughness of approximately $\pm$2 nm and very few spikes (right image). We refer to this ITO texture as ``Smooth". Statistical analyses of these images (Fig. \ref{Fig_SI_ITO_Spikes_Analysis}b) indicate that ``Spiky" ITO has on average greater than 3 distinct protrusions (``spikes") per square micrometer and several million ITO protrusions are expected in the area of our 12 mm$^2$ devices. ``Smooth" ITO has fewer than 0.1 distinct spikes per square micrometer, and they are on average shorter. The approximate proportion of exposed ITO was calculated using a simple ``fluid" model by treating the ITO spikes as circular cylinders and the PTAA as a perfect (zero viscosity) fluid (Fig. \ref{Fig_SI_ITO_Spikes_Analysis}c). Fig. \ref{Fig_SI_ITO_Spikes_Analysis}d shows the estimated portion of exposed ITO as a function of PTAA thickness using that model. \\

\begin{figure}[h!]
    \centering
    \includegraphics[width=3.5in]{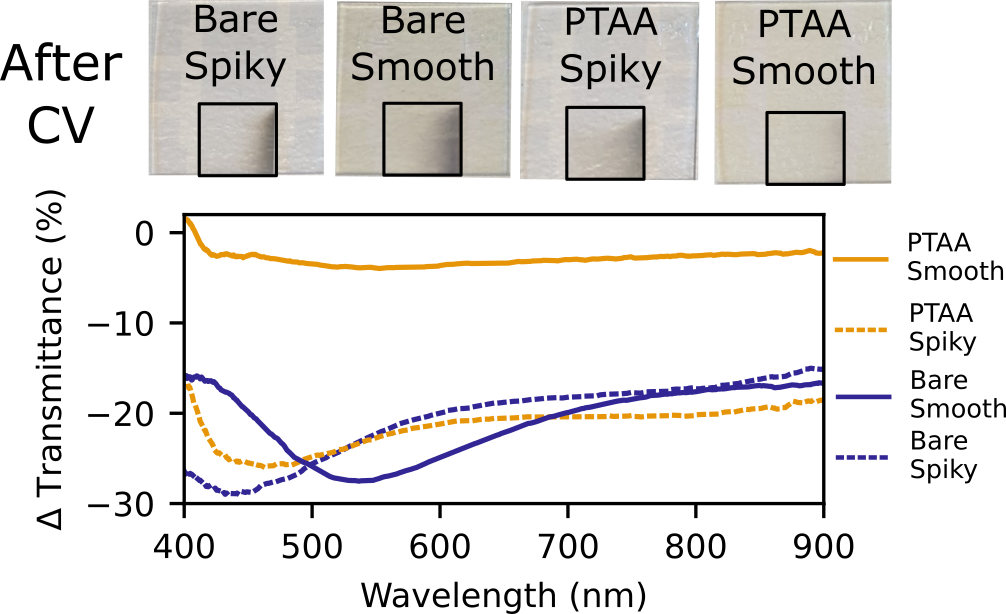}
	\caption{
     Optical images (top) and optical transmittance spectra (bottom) through ITO substrates after a single CV cycle from 0 V to $-1.55$ V and back to 0 V (relative to an Ag/AgCl reference electrode), in a solution of methylammonium iodide in dimethylformamide. Black squares in the optical images represent the areas contacting the solution during the CV cycle.
    }
\label{Fig_Cyclic_Voltammetry}
\end{figure}

 We assessed the ability of a thin solution-processed HTL (PTAA) to cover textured ITO by quantifying the degree of irreversible electrochemical reactions that occur between ITO and a solution of methylammonium iodide in dimethylformamide. Spiky and smooth ITO were either left bare or were coated with about 8 nm of PTAA. Samples were subjected to a single cyclic voltammetry (CV) cycle from 0 V to $-1.55$ V and back to 0 V (relative to an Ag/AgCl reference electrode). Exposed ITO becomes optically dark during this process, losing about 20\% to 30\% of its transparency, as seen in optical images and in optical transmittance spectra (Fig. \ref{Fig_Cyclic_Voltammetry}). ITO darkening arises from electrochemical reduction of In$^{3+}$ in the ITO (In$_2$O$_3$) to metallic indium and the formation of sub-oxides at the surface \cite{kerner_electrochemical_2019, minenkov_monitoring_2024}. Smooth ITO covered with PTAA showed substantially less darkening, losing only about 3\% of its transparency. Spiky ITO covered with PTAA lost about 20\% of its transparency, similar to bare ITO. The PTAA inhibits indium reactions by blocking electron transfer between the ITO and the electrolyte. PTAA also likely inhibits electrochemistry by keeping the ITO chemically separated from ions (including methylammonium and iodide) in the solution \cite{kerner_electrochemical_2019, minenkov_monitoring_2024}. 

\end{document}